\documentclass[english,prd,showpacs,showkeys,preprintnumbers,notitlepage,nofootinbib]{revtex4}
\usepackage[letterpaper]{geometry}
\usepackage{float}
\usepackage{amsmath}
\usepackage{amssymb}
\usepackage{graphicx}
\usepackage{esint}
\usepackage[unicode=true,pdfusetitle,
 bookmarks=true,bookmarksnumbered=false,bookmarksopen=false,
 breaklinks=false,pdfborder={0 0 1},backref=false,colorlinks=false]
 {hyperref}

\makeatletter

\providecommand{\tabularnewline}{\\}

\def\Pom{{ I\!\!P}}

\makeatother

\begin{document}

\title{Flavor structure of generalized parton distributions from neutrino
experiments}

\author{B.~Z.~Kopeliovich, Iv\'{a}n~Schmidt and M.~Siddikov}

\address{Departamento de F\'{i}sica, Instituto de Estudios Avanzados en Ciencias
e Ingenier\'{i}a, y Centro Cient\'{i}fico - Tecnológico de Valpara\'{i}so,
Universidad T\'{e}cnica Federico Santa Mar\'{i}a, Casilla 110-V, Valpara\'{i}so,
Chile}

\preprint{USM-TH-307}
\begin{abstract}
The analysis of deeply virtual meson production is extended to neutrino-production
of the pseudo-Goldstone mesons ($\pi,\, K,\,\eta$) on nucleons, with
the flavor content of the recoil baryon either remaining intact, or
changing to a hyperon from the $SU(3)$ octet. We rely on the $SU(3)$
 relations and express all the cross-sections in terms of the proton
generalized parton distributions (GPDs). The corresponding amplitudes
are calculated at the leading twist level and in the leading order
in $\alpha_{s}$, using a phenomenological model of GPDs. We provide
a computational code, which can be used for evaluation of the cross-sections
employing various GPD models. We conclude that these processes can
be studied in the experiment \textsc{Minerva} at FERMILAB, which could
supplement the measurements at JLAB helping to extract the GPD
flavor structure from data. 
\end{abstract}

\pacs{13.15.+g,13.85.-t}

\keywords{Single pion production, generalized parton distributions, neutrino-hadron
interactions}

\maketitle

\section{Introduction}

During the last decade the notion of generalized parton distributions
(GPDs) became a standard theoretical tool to describe the nonperturbative
structure of the hadronic target. These new objects, being an special case
of the general Wigner distributions, contain rich information about
the nonperturbative dynamics of the target structure, such as form
factors, ordinary parton distribution functions (PDF), fractions of
the spin carried by each parton, etc. (see e.g. recent reviews in~\cite{Kumericki:2009uq,Boffi:2007yc,Belitsky:2005qn,Diehl:2004cx,Diehl:2003ny,Sabatie:2012pe}).
In hard exclusive reactions in the Bjorken kinematics, due to collinear
factorization~\cite{Ji:1998xh,Collins:1998be} the amplitude of many
processes may be represented as a convolution of the process-dependent
perturbative coefficient functions with target-dependent GPDs. While
a model-independent deconvolution and extraction of GPDs from data
is in general impossible%
\footnote{An exception is the process of double deeply virtual Compton scattering
discussed in~\cite{Guidal:2002kt,Belitsky:2002tf,Belitsky:2003fj},
however vanishingly small cross-sections make it unreachable at modern
accelerators.%
}, nevertheless data help to constrain the available models of GPDs.

Currently the main source of information about GPDs are the electron-proton
measurements done at JLAB and HERA, in particular deeply virtual Compton
scattering (DVCS) and deeply virtual meson production (DVMP)~\cite{Mueller:1998fv,Ji:1996nm,Ji:1998pc,Radyushkin:1996nd,Radyushkin:1997ki,Radyushkin:2000uy,Ji:1998xh,Collins:1998be,Collins:1996fb,Brodsky:1994kf,Goeke:2001tz,Diehl:2000xz,Belitsky:2001ns,Diehl:2003ny,Belitsky:2005qn,Kubarovsky:2011zz}.
A planned CLAS12 upgrade at JLAB will help to improve our understanding
of the GPDs~\cite{Kubarovsky:2011zz}.

Having only data on DVCS one cannot single out the flavor structure
of GPDs. The process of DVMP potentially is able to disentangle the
flavor structure of the GPDs, since different mesons are sensitive
to different GPD flavor combinations~\cite{Vanderhaeghen:1998uc,Mankiewicz:1998kg}.
However, the practical realization of this program suffers from large
uncertainties. In the HERA kinematics ($x_{Bj}\lesssim10^{-2}$), one
is close to the saturation regime, where gluons dominate, and as was
discussed in~\cite{Ivanov:2007je}, NLO corrections in this kinematical range
become large due to BFKL-type logarithms. As a consequence, for the description
of exclusive processes one should use models where the saturation
is built-in~\cite{GolecBiernat:1998js,Hufner:2000jb,Kopeliovich:1999am,Kowalski:2003hm}.
At the same time, in the JLAB kinematics, the range of $Q^{2}$ is
quite restricted, and the GPDs extracted from DVCS may be essentially
contaminated by higher-twist effects. In case of the DVMP, an
additional uncertainty comes from the distribution amplitudes (DA)
of the produced mesons: while there is a lot of models, only the DAs
for $\pi$ and $\eta$ were confronted with data~\cite{Gronberg:1997fj,Aubert:2009mc,Uehara:2012ag}
(see also the recent review in~\cite{Brodsky:2011xx,Brodsky:2011yv}).
For heavier mesons ($\rho,\omega,\phi$), the DAs are completely unknown,
because their partonic structure is controlled by confinement,
rather than by the chiral symmetry as for Goldstone mesons. In the general
case it is not even known if the corresponding DAs should vanish at
the endpoints of the fractional light-cone momentum distribution,
as is required by collinear factorization, or if the amplitude gets a
contribution from transverse degrees of freedom.

From this point of view, consistency checks of GPD extraction from
JLAB data, especially of their flavor structure, are important. Neutrino
experiments present a powerful tool, which could be used for this
purpose. Up to recently, the high-precision exclusive neutrino-hadron
differential cross-sections were available only in the low-energy
region, where the physics is described by $s$-channel resonances~\cite{AlvarezRuso:2012fc,Alarcon:2011kh,Praet:2008yn,Paschos:2011ye,RafiAlam:2010kf}.
In the high-energy regime, due to the smallness of the cross-sections
and vanishingly small luminosities in the tails of the neutrino spectra,
the experimental data have been available so far only either for inclusive
or for integrated (total) exclusive cross-sections. The situation
is going to change next year, when the high-intensity NuMI beam at
Fermilab will switch on to the so-called middle-energy (ME) regime with
an average neutrino energy of about $6$~GeV. In this setup the \textsc{Minerva}
experiment\textsc{~\cite{Drakoulakos:2004gn}} should be able to
probe the quark flavor structure of the targets. Potentially, \textsc{NuMI}
neutrino beam may reach energies up to 20 GeV, without essential loss
of luminosity. 
Even higher luminosities in multi-GeV regime can be achieved at the
planned Muon Collider/Neutrino Factory~\cite{Gallardo:1996aa,Ankenbrandt:1999as,Alsharoa:2002wu}.

One can access the GPD flavor structure in neutrino interactions by
studying the same processes as in \emph{ep} collisions and employing
the difference of the weak and electromagnetic couplings for the vector
current. An example of such a process is the weak DVCS discussed in~\cite{Psaker:2006gj}.
However, the weak DVCS alone is not sufficient to constrain the flavor
structure. Moreover, the typical magnitude of the cross-sections for
such processes is tiny, of the order $\sim10^{-44}cm^{2}/GeV^{4}$.

The $\nu$DVMP measurements with neutrino and antineutrino beams are
complementary to the electromagnetic DVCS. In the axial channel, due
to the chiral symmetry breaking we have an octet of pseudo-Goldstone bosons
which act as a natural probe for the flavor content. Due to the $V-A$
structure of the charged current,  in $\nu$ DVMP one can access simultaneously
the unpolarized GPDs, $H,E$, and the helicity flip GPDs, $\tilde{H}$
and $\tilde{E}$. Besides, important information on flavor structure
can be obtained by studying the transitional GPDs in the processes
with nucleon to hyperon transitions. As was discussed in~\cite{Frankfurt:1999fp},
due to $SU(3)$ flavor symmetry, these GPDs can be related to the
ordinary diagonal GPDs in the proton.

The paper is organized as follows. In Section~\ref{sec:DVMP_Xsec}
we evaluate the Goldstone meson production by neutrinos on nucleon targets. 
The main result of this section is Table~\ref{tab:DVMP_amps}
and Eqns~\ref{eq:XSec}-\ref{eq:T2_unp}. In Section~\ref{sec:Parametrizations},
for the sake of completeness we sketch the properties of the GPD parametrization
which will be used for evaluations. In Section~\ref{sec:Results}
we present numerical results and make conclusions.

\section{Cross-section of the $\nu$DVMP process}

\label{sec:DVMP_Xsec} 
\begin{figure}[htp]
\includegraphics[scale=0.4]{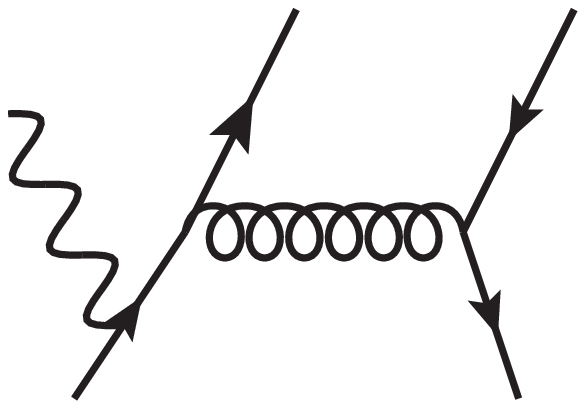}\includegraphics[scale=0.4]{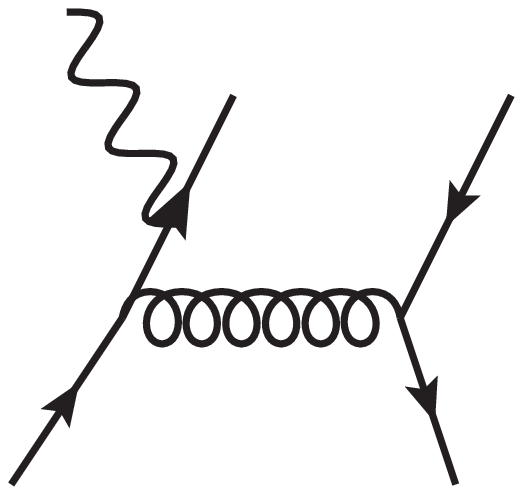}\includegraphics[scale=0.4]{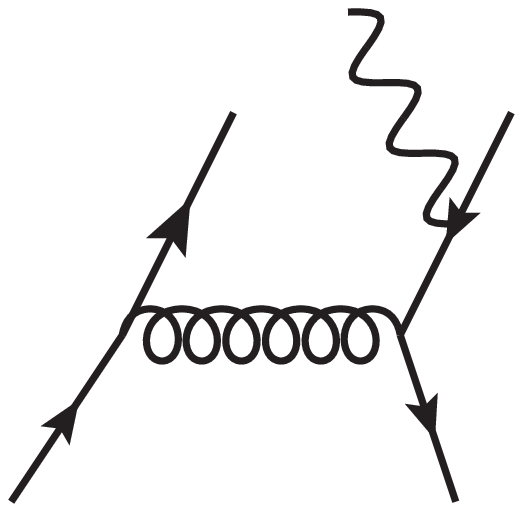}\includegraphics[scale=0.4]{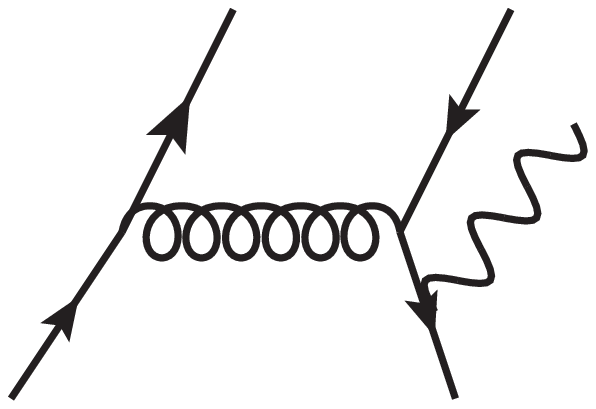}
\caption{\label{fig:DVMPLT}Leading-order and leading twist contributions to the
DVMP hard coefficient functions.}
\end{figure}

The DVMP process in the vector channel has
been studied in~\cite{Vanderhaeghen:1998uc,Mankiewicz:1998kg}. In
the leading order in $\alpha_{s}$, the hard coefficient function gets
contributions from the diagrams shown in Figure~\ref{fig:DVMPLT}.

Evaluation of these diagrams is straightforward and yields for the
amplitude of the process, 
\begin{eqnarray}
T_{M} & = & \frac{8\pi i}{9}\frac{\alpha_{s}f_{M}}{Q}\left(\int dz\frac{\phi_{M}(z)}{z}\right)\sum_{\Gamma}\mathcal{H}_{M}^{\Gamma}\bar{N}\left(p_{2}\right)\Gamma N\left(p_{1}\right),\label{eq:T_def}
\end{eqnarray}
where $N(p),\,\bar{N}(p)$ are the spinors of the initial/final state
baryon, $\phi_{M}(z)$ is the normalized to unity distribution amplitude of the produced
meson, $f_{M}$ is the decay constant of the corresponding meson,
$\sum_{\Gamma}H_{M}^{\Gamma}\bar{N}\left(p_{2}\right)\Gamma N\left(p_{1}\right)$
is a symbolic notation for summation of all leading twist GPDs contributions
(defined below), and $\mathcal{H}_{M}^{\Gamma}$ are the convolutions
of the GPDs $H_{\Gamma}$ of the target with the proper coefficient
function. Currently, the amplitude of the DVMP is known up to NLO
accuracy~\cite{Ivanov:2004zv,Diehl:2007hd}. Extension of the analysis
of~\cite{Vanderhaeghen:1998uc,Mankiewicz:1998kg} to neutrinos is
straightforward. In contrast to electroproduction, due to $V-A$ structure,
the amplitudes acquire contributions from both the unpolarized and
helicity flip GPDs.

In the leading twist, four GPDs, $H,\, E,\,\tilde{H}$ and $\tilde{E}$
contribute to this process. They are defined as 
\begin{eqnarray}
\frac{\bar{P}^{+}}{2\pi}\int dz\, e^{ix\bar{P}^{+}z}\left\langle B\left(p_{2}\right)\left|\bar{\psi}_{q'}\left(-\frac{z}{2}\right)\gamma_{+}\psi_{q}\left(\frac{z}{2}\right)\right|A\left(p_{1}\right)\right\rangle  & = & \left(H_{q}\left(x,\xi,t\right)\bar{N}\left(p_{2}\right)\gamma_{+}N\left(p_{1}\right)\right.\label{eq:H_def}\\
 &  & \left.+\frac{\Delta_{k}}{2m_{N}}E_{q}\left(x,\xi,t\right)\bar{N}\left(p_{2}\right)i\sigma_{+k}N\left(p_{1}\right)\right)\nonumber \\
\frac{\bar{P}^{+}}{2\pi}\int dz\, e^{ix\bar{P}^{+}z}\left\langle B\left(p_{2}\right)\left|\bar{\psi}_{q'}\left(-\frac{z}{2}\right)\gamma_{+}\gamma_{5}\psi_{q}\left(\frac{z}{2}\right)\right|A\left(p_{1}\right)\right\rangle  & = & \left(\tilde{H}_{q}\left(x,\xi,t\right)\bar{N}\left(p_{2}\right)\gamma_{+}\gamma_{5}N\left(p_{1}\right)\right.\label{eq:Htilde_def}\\
 &  & \left.+\frac{\Delta_{+}}{2m_{N}}\tilde{E}_{q}\left(x,\xi,t\right)\bar{N}\left(p_{2}\right)N\left(p_{1}\right)\right),\nonumber 
\end{eqnarray}
where $\bar{P}=p_{1}+p_{2}$, $\Delta=p_{2}-p_{1}$ and $\xi=-\Delta^{+}/2\bar{P}^{+}\approx x_{Bj}/(2-x_{Bj})$
(see e.g.~\cite{Goeke:2001tz} for details of kinematics). In the general
case, when $A\not=B$, in the right-hand side (r.h.s.) of Eqs.~(\ref{eq:H_def}),
(\ref{eq:Htilde_def}) there might be extra structures which are forbidden
by $T$-parity in the case of $A=B$~\cite{Goeke:2001tz}. In what
follows we assume that the initial state $A$ is either a proton or
a neutron, and $B$ belongs to the same lowest $SU(3)$ octet of baryons.
In this case, all such terms are parametrically suppressed by the
current quark mass $m_{q}$ and vanish in the limit of exact $SU(3)$,
so we will disregard them. Since in neutrino experiments the target
cannot be polarized due to its large size, it makes no sense to discuss
contributions of the transversity GPDs $H_{T},\, E_{T},\,\tilde{H}_{T},\,\tilde{E}_{T}$.
Also, in this paper we ignore the contributions of gluons, because
in the current and forthcoming neutrino experiments the region of
small $x_{Bj}\ll1$ is not achievable, so the amplitude~(\ref{eq:T_def})
simplifies to 
\begin{eqnarray}
T_{M} & = & \frac{8\pi i}{9}\frac{\alpha_{s}f_{M}}{Q}\left(\int dz\frac{\phi_{M}(z)}{z}\right)\left[\left(\tilde{\mathcal{H}}_{M}\bar{N}\left(p_{2}\right)\gamma_{+}\gamma_{5}N\left(p_{1}\right)+\frac{\Delta_{+}}{2m_{N}}\tilde{\mathcal{E}}_{M}\bar{N}\left(p_{2}\right)\gamma_{5}N\left(p_{1}\right)\right)\right.\label{eq:T_LT}\\
 & + & \left.\left(\mathcal{H}_{M}\bar{N}\left(p_{2}\right)\gamma_{+}N\left(p_{1}\right)+\frac{\Delta_{k}}{2m_{N}}\mathcal{E}_{M}\bar{N}\left(p_{2}\right)i\sigma_{+k}N\left(p_{1}\right)\right)\right],\nonumber 
\end{eqnarray}

In table~\ref{tab:DVMP_amps} the corresponding amplitudes are listed
for each final state $M$. It is restricted to the cases of either
protons or neutrons in the initial state, and only baryons from the
lowest lying octet in the final state. We used ordinary $SU(3)$ relations~\cite{Frankfurt:1999fp}
to relate the transitional GPDs $\left\langle Y\left|\hat{\mathcal{O}}_{q,q'}\right|p\right\rangle $
to the proton GPDs $\left\langle p\left|\hat{\mathcal{O}}_{q,q}\right|p\right\rangle $.
As was mentioned in~\cite{Frankfurt:1999fp}, these relations for
the GPD $\tilde{E}$ can be broken due to the different masses of pion
and kaon in the $t$-channel. Also, the $SU(3)$ relations can be
inaccurate at small-$x_{Bj}$ (high energy), due to different intercepts
of the $\pi/\rho$ and $K/K^{*}$ Regge trajectories~%
\footnote{Similar results may be obtained in the framework of the dipole model~\cite{GolecBiernat:1998js,Hufner:2000jb,Kopeliovich:1999am,Kowalski:2003hm},
which is valid for very small $x\lesssim10^{-2}$: The amplitude gets
a substantial contribution from the endpoint region with $\alpha$ or
$\bar{\alpha}\sim m_{q}^{2}/Q^{2}$. This asymmetry in the $\alpha$
contribution depends strongly on the quark mass and obviously breaks
the $SU(3)$ symmetry.%
}~\cite{Guidal:1997hy,Li:2004gu}. Besides, as was discussed in~\cite{Ivanov:2007je},
in the small-$x_{Bj}$ regime NLO corrections become large due to BFKL-type
logarithms, and a lot of care is needed to make a systematic resummation
and avoid double counting. For this reason, in what follows we restrict
our consideration to the moderate energy range ($x_{Bj}\gtrsim0.1$).
For a neutron target, in the right panel of table~\ref{tab:DVMP_amps}
we flipped $H_{u/n}\to H_{d/p}$,~$H_{d/n}\to H_{u/p}$, so all the
GPDs are given for a proton target. The corresponding constants
$g_{f}$ should be understood as neutral current couplings $g_{A}^{f}$
and $g_{V}^{f}$ for the DVMP form factors $\mathcal{H}$, $\mathcal{E}$,
and $\tilde{\mathcal{H}}$, $\tilde{\mathcal{E}}$ respectively. Electroproduction
data~\cite{Vanderhaeghen:1998uc,Mankiewicz:1998kg} correspond to
the mere change of the charges, $g_{A}\to 0,\, g_{V}^{f}\to e_{f}$.

\begin{table}[h]
\caption{\label{tab:DVMP_amps}List of the DVMP amplitudes $\mathcal{H}_{M},\,\mathcal{E}_{M},\,\tilde{\mathcal{H}}_{M},\,\tilde{\mathcal{E}}_{M}$
for different final states. For a neutron target, in the r.h.s.
we flipped $H_{u/n}\to H_{d/p}$,~$H_{d/n}\to H_{u/p}$, so all the
GPDs are given for a proton target. To get $\mathcal{E}$, $\tilde{\mathcal{H}}$,
$\tilde{\mathcal{E}}$, replace $H$ with $E$, $\tilde{H}$, $\tilde{E}$
respectively. The corresponding constants $g_{f}$ should be understood
as neutral current couplings $g_{A}^{f}$ for the DVMP form factors
$\mathcal{H}$, $\mathcal{E}$, and as $g_{V}^{f}$ for $\tilde{\mathcal{H}}$,
$\tilde{\mathcal{E}}$. $V_{ij}$ are the CKM matrix elements. $c_{\pm}$
is a shorthand notation $c_{\pm}(x,\xi)=1/(x\pm\xi\mp i0)$ for the
leading order coefficient function. NLO corrections to the coefficient
functions may be found in~\cite{Ivanov:2004zv,Diehl:2007hd}. For
the sake of brevity, we did not show the arguments $(x,\xi,t,Q)$
for all GPDs and omitted the integral over the quark light-cone fraction
$\int dx$ everywhere. }

\global\long\def\arraystretch{1.5}
\begin{tabular}{|c|c|c|c|c|c|c|}
\cline{1-3} \cline{5-7} 
Process  & type  & $\mathcal{H}_{M}$  &  & Process  & type  & $\mathcal{H}_{M}$\tabularnewline
\cline{1-3} \cline{5-7} 
$\nu\, p\to\mu^{-}\pi^{+}p$  & CC  & $V_{ud}\left(H_{d}c_{-}+H_{u}c_{+}\right)$  &  & $\nu\, n\to\mu^{-}\pi^{+}n$  & CC  & $V_{ud}\left(H_{u}c_{-}+H_{d}c_{+}\right)$\tabularnewline
\cline{1-3} \cline{5-7} 
$\bar{\nu}\, p\to\mu^{+}\pi^{-}p$  & CC  & $V_{ud}\left(H_{u}c_{-}+H_{d}c_{+}\right)$  &  & $\bar{\nu}\, n\to\mu^{+}\pi^{-}n$  & CC  & $V_{ud}\left(H_{d}c_{-}+H_{u}c_{+}\right)$\tabularnewline
\cline{1-3} \cline{5-7} 
$\bar{\nu\,}p\to\mu^{+}\pi^{0}n$  & CC  & $V_{ud}\left(H_{u}-H_{d}\right)\left(c_{+}-c_{-}\right)/\sqrt{2}$  &  & $\nu\, n\to\mu^{-}\pi^{0}p$  & CC  & $V_{ud}\left(H_{u}-H_{d}\right)\left(c_{-}-c_{+}\right)/\sqrt{2}$\tabularnewline
\cline{1-3} \cline{5-7} 
$\nu\, p\to\nu\,\pi^{+}n$  & NC  & $\left(H_{u}-H_{d}\right)\left(g_{u}c_{-}+g_{d}c_{+}\right)$  &  & $\nu\, n\to\nu\,\pi^{-}p$  & NC  & $\left(H_{u}-H_{d}\right)\left(g_{d}c_{-}+g_{u}c_{+}\right)$\tabularnewline
\cline{1-3} \cline{5-7} 
$\nu\, p\to\nu\,\pi^{0}p$  & NC  & $\left(g_{u}H_{u}-g_{d}H_{d}\right)\left(c_{-}+c_{+}\right)/\sqrt{2}$  &  & $\nu\, n\to\nu\,\pi^{0}n$  & NC  & $\left(g_{u}H_{d}-g_{d}H_{u}\right)\left(c_{-}+c_{+}\right)/\sqrt{2}$\tabularnewline
\cline{1-3} \cline{5-7} 
$\bar{\nu\,}p\to\mu^{+}\pi^{-}\Sigma_{+}$  & CC  & $-V_{us}\left(H_{d}-H_{s}\right)c_{+}$  &  & $\bar{\nu\,}n\to\mu^{+}\pi^{-}\Lambda$  & CC  & $-V_{us}\left(2H_{d}-H_{u}-H_{s}\right)c_{+}/\sqrt{6}$\tabularnewline
\cline{1-3} \cline{5-7} 
$\bar{\nu\,}p\to\mu^{+}\pi^{0}\Sigma_{0}$  & CC  & $V_{us}\left(H_{d}-H_{s}\right)c_{+}/2$  &  & $\bar{\nu\,}n\to\mu^{+}\pi^{-}\Sigma_{0}$  & CC  & $-V_{us}\left(H_{u}-H_{s}\right)c_{+}/\sqrt{2}$\tabularnewline
\cline{1-3} \cline{5-7} 
$\bar{\nu\,}p\to\mu^{+}\pi^{0}\Lambda$  & CC  & $V_{us}\left(2H_{u}-H_{d}-H_{s}\right)c_{+}/2\sqrt{3}$  &  & $\bar{\nu\,}n\to\mu^{+}\pi^{0}\Sigma^{-}$  & CC  & $V_{us}\left(H_{u}-H_{s}\right)c_{+}/\sqrt{2}$\tabularnewline
\cline{1-3} \cline{5-7} 
\multicolumn{1}{c}{} & \multicolumn{1}{c}{} & \multicolumn{1}{c}{} & \multicolumn{1}{c}{} & \multicolumn{1}{c}{} & \multicolumn{1}{c}{} & \multicolumn{1}{c}{}\tabularnewline
\cline{1-3} \cline{5-7} 
$\nu\, p\to\mu^{-}K^{+}p$  & CC  & $V_{us}\left(c_{+}H_{u}+c_{-}H_{s}\right)$  &  & $\nu\, n\to\mu^{-}K^{+}n$  & CC  & $V_{us}\left(c_{+}H_{d}+c_{-}H_{s}\right)$\tabularnewline
\cline{1-3} \cline{5-7} 
$\bar{\nu}\, p\to\mu^{+}K^{-}p$  & CC  & $V_{us}\left(H_{u}c_{-}+H_{s}c_{+}\right)$  &  & $\bar{\nu}\, n\to\mu^{+}K^{-}n$  & CC  & $V_{us}\left(H_{d}c_{-}+H_{s}c_{+}\right)$\tabularnewline
\cline{1-3} \cline{5-7} 
$\bar{\nu}\, p\to\mu^{+}K^{0}\Sigma_{0}$  & CC  & $-V_{ud}\left(H_{d}-H_{s}\right)c_{-}/\sqrt{2}$  &  & $\bar{\nu}\, n\to\mu^{+}K^{0}\Sigma^{-}$  & CC  & $-V_{ud}\left(H_{u}-H_{s}\right)c_{-}$\tabularnewline
\cline{1-3} \cline{5-7} 
$\bar{\nu}\, p\to\mu^{+}K^{0}\Lambda$  & CC  & $-V_{ud}\left(2H_{u}-H_{d}-H_{s}\right)c_{-}/\sqrt{6}$  &  & $\nu\, n\to\nu K^{0}\Lambda$  & NC  & $-g_{d}\left(2H_{d}-H_{u}-H_{s}\right)\left(c_{-}+c_{+}\right)/\sqrt{6}$\tabularnewline
\cline{1-3} \cline{5-7} 
$\bar{\nu}\, p\to\mu^{+}\bar{K}^{0}n$  & CC  & $-V_{us}\left(H_{u}-H_{d}\right)c_{-}$  &  & $\nu\, n\to\nu K^{0}\Sigma_{0}$  & NC  & $-g_{d}\left(H_{u}-H_{s}\right)\left(c_{-}+c_{+}\right)/\sqrt{2}$\tabularnewline
\cline{1-3} \cline{5-7} 
$\nu\, p\to\mu^{-}K^{+}\Sigma^{+}$  & CC  & $-V_{ud}\left(H_{d}-H_{s}\right)c_{-}$  &  & $\nu\, n\to\mu^{-}K^{+}\Sigma^{0}$  & CC  & $-V_{ud}\left(H_{u}-H_{s}\right)c_{-}/\sqrt{2}$\tabularnewline
\cline{1-3} \cline{5-7} 
$\nu\, p\to\nu\, K^{+}\Lambda$  & NC  & $-\left(2H_{u}-H_{d}-H_{s}\right)\left(g_{u}c_{-}+g_{d}c_{+}\right)/\sqrt{6}$  &  & $\nu\, n\to\mu^{-}\, K^{+}\Lambda$  & CC  & $-V_{ud}\left(2H_{d}-H_{u}-H_{s}\right)c_{-}/\sqrt{6}$\tabularnewline
\cline{1-3} \cline{5-7} 
$\nu\, p\to\nu\, K^{+}\Sigma_{0}$  & NC  & $\left(H_{d}-H_{s}\right)\left(g_{u}c_{-}+g_{d}c_{+}\right)/\sqrt{2}$  &  & $\nu\, n\to\mu^{-}K^{0}p$  & CC  & $-V_{us}\left(H_{d}-H_{u}\right)c_{+}$\tabularnewline
\cline{1-3} \cline{5-7} 
$\nu\, p\to\nu\, K^{0}\Sigma^{+}$  & NC  & $-g_{d}\left(H_{d}-H_{s}\right)\left(c_{-}+c_{+}\right)$  &  & $\nu\, n\to\nu\, K^{+}\Sigma^{-}$  & NC  & $-\left(H_{u}-H_{s}\right)\left(g_{u}c_{-}+g_{d}c_{+}\right)$\tabularnewline
\cline{1-3} \cline{5-7} 
\multicolumn{1}{c}{} & \multicolumn{1}{c}{} & \multicolumn{1}{c}{} & \multicolumn{1}{c}{} & \multicolumn{1}{c}{} & \multicolumn{1}{c}{} & \multicolumn{1}{c}{}\tabularnewline
\cline{1-3} \cline{5-7} 
$\nu\, p\to\nu\,\eta\, p$  & NC  & $\left(g_{u}H_{u}+g_{d}H_{d}-2g_{d}H_{s}\right)\left(c_{-}+c_{+}\right)/\sqrt{6}$  &  & $\nu\, n\to\nu\,\eta\, n$  & NC  & $\left(g_{u}H_{d}+g_{d}H_{u}-2g_{d}H_{s}\right)\left(c_{-}+c_{+}\right)/\sqrt{6}$\tabularnewline
\cline{1-3} \cline{5-7} 
$\bar{\nu}\, p\to\mu^{+}\,\eta\, n$  & CC  & $V_{ud}\left(H_{u}-H_{d}\right)\left(c_{-}+c_{+}\right)/\sqrt{6}$  &  & $\bar{\nu}\, n\to\mu^{+}\,\eta\,\Sigma^{-}$  & CC  & $V_{us}\left(H_{u}-H_{s}\right)\left(2\, c_{-}-c_{+}\right)/\sqrt{6}$\tabularnewline
\cline{1-3} \cline{5-7} 
$\bar{\nu}\, p\to\mu^{+}\,\eta\,\Sigma_{0}$  & CC  & $V_{us}\left(H_{u}-H_{d}\right)\left(c_{+}-2\, c_{-}\right)/2\sqrt{3}$  &  & $\nu\, n\to\mu^{-}\,\eta\, p$  & CC  & $V_{ud}\left(H_{u}-H_{d}\right)\left(c_{-}+c_{+}\right)/\sqrt{6}$\tabularnewline
\cline{1-3} \cline{5-7} 
$\bar{\nu}\, p\to\mu^{+}\,\eta\,\Lambda$  & CC  & $V_{us}\left(2H_{u}-H_{d}-H_{s}\right)\left(c_{+}-2\, c_{-}\right)/6$  &  &  &  & \tabularnewline
\cline{1-3} \cline{5-7} 
\end{tabular}
\end{table}

While in the table we listed, for the purpose of reference, all 41 amplitudes,
only 12 of them are independent due to the $SU(3)$ symmetry%
\footnote{This follows from number of irreps in \textbf{8$\times$8=1+8+8+10+10{*}+27}\textbf{\emph{
}}for a given\textbf{\emph{ $J^{P}$}}%
}. This agrees with the fact that all the amplitudes are linear combinations
of 6 functions $\int dx\, H_{u,d,s}\left(x,\xi,t\right)c_{\pm}(x,\xi)$
and 6 functions $\int dx\,\tilde{H}_{u,d,s}\left(x,\xi,t\right)c_{\pm}(x,\xi)$
for the axial and vector channels respectively%
\footnote{As one can see from~\ref{eq:T2_unp}, the GPDs $E,\,\tilde{E}$ always
contribute in bilinear combinations $H^{*}H-E^{*}E$ with the same
coefficients as the GPDs $H,\tilde{H}$, so this does not change the total
count of independent cross-sections%
}. This implies a large number of relations between different cross-sections,
some of which are obvious. For example, comparing different elements
of the Table~\ref{tab:DVMP_amps}, we may get%
\footnote{For example, the cross-sections of $\nu\, p\to\mu^{-}\pi^{+}p$ and
$\bar{\nu}\, n\to\mu^{+}\pi^{-}n$ are equal because the amplitudes
of subprocesses $W^{+}p\to\pi^{+}p$ and $W^{-}n\to\pi^{-}n$ coincide
due to isospin symmetry ($I=3/2$ state).%
}: 
\begin{table}[H]
\centering{}\global\long\def\arraystretch{1.5}
\begin{tabular}{ccc}
$d\sigma_{\nu\, p\to\mu^{-}\pi^{+}p}=d\sigma_{\bar{\nu}\, n\to\mu^{+}\pi^{-}n},\quad$  & $d\sigma_{\bar{\nu\,}p\to\mu^{+}\pi^{0}n}=d\sigma_{\nu\, n\to\mu^{-}\pi^{0}p},\quad$  & $d\sigma_{\nu\, n\to\mu^{-}\pi^{+}n}=d\sigma_{\bar{\nu}\, p\to\mu^{+}\pi^{-}p},$\tabularnewline
$d\sigma_{\nu\, p\to\mu^{-}K^{+}\Sigma^{+}}=2\, d\sigma_{\bar{\nu}\, p\to\mu^{+}K^{0}\Sigma_{0}},$  & $d\sigma_{\bar{\nu}\, n\to\mu^{+}K^{0}\Sigma^{-}}=2\, d\sigma_{\nu\, n\to\mu^{-}K^{+}\Sigma^{0}},$  & $d\sigma_{\bar{\nu}\, p\to\mu^{+}\,\eta\, n}=d\sigma_{\nu\, n\to\mu^{-}\,\eta\, p},$\tabularnewline
$d\sigma_{\bar{\nu\,}n\to\mu^{+}\pi^{-}\Sigma_{0}}=d\sigma_{\bar{\nu\,}n\to\mu^{+}\pi^{0}\Sigma^{-}}$  & $d\sigma_{\bar{\nu\,}p\to\mu^{+}\pi^{-}\Sigma_{+}}=4\, d\sigma_{\bar{\nu\,}p\to\mu^{+}\pi^{0}\Sigma_{0}}$  & $d\sigma_{\bar{\nu\,}n\to\mu^{+}\pi^{-}\Sigma_{0}}=d\sigma_{\bar{\nu\,}n\to\mu^{+}\pi^{0}\Sigma^{-}}$\tabularnewline
\end{tabular}
\end{table}

Other relations can be extracted using the identity 
\begin{equation}
\left|A+B\right|^{2}+\left|A-B\right|^{2}=2\left(\left|A\right|^{2}+\left|B\right|^{2}\right).\label{eq:_idty}
\end{equation}
For example, fixing 
\[
A=\left(H_{u}-H_{d}\right)c_{-},\quad B=\left(H_{u}-H_{d}\right)c_{+},
\]
we arrive at the relation between the Cabibbo-suppressed and Cabibbo-allowed
cross-sections 
\[
\left(d\sigma_{\bar{\nu}\, p\to\mu^{+}\bar{K}^{0}n}+d\sigma_{\nu\, n\to\mu^{-}K^{0}p}\right)=\left|\frac{V_{us}}{V_{ud}}\right|^{2}\left(d\sigma_{\nu\, n\to\mu^{-}\pi^{0}p}+3\, d\sigma_{\bar{\nu}\, p\to\mu^{+}\,\eta\, n}\right).
\]

The corresponding neutrino cross-sections for charged and neutral
currents read, 
\begin{eqnarray}
\frac{d^{3}\sigma_{CC}}{dt\, d\ln x_{Bj}\,\, dQ^{2}} & = & \frac{G_{F}^{2}x_{Bj}^{2}\left(1-y-\frac{m_{N}^{2}x^{2}y^{2}}{Q^{2}}\right)}{32\pi^{3}Q^{2}\left(1+Q^2/M_W^2\right)^2\left(1+\frac{4m_{N}^{2}x_{Bj}^{2}}{Q^{2}}\right)^{3/2}}\left|T_{M}\right|^{2},\label{eq:XSec}\\
\frac{d^{3}\sigma_{NC}}{dt\, d\ln x_{Bj}\,\, dQ^{2}} & = & \frac{G_{F}^{2}x_{Bj}^{2}\left(1-y-\frac{m_{N}^{2}x^{2}y^{2}}{Q^{2}}\right)}{32\pi^{3}\cos^{4}\theta_{W}Q^{2}\left(1+Q^2/M_Z^2\right)^2\left(1+\frac{4m_{N}^{2}x_{Bj}^{2}}{Q^{2}}\right)^{3/2}}\left|T_{M}\right|^{2}.
\end{eqnarray}

In analogy to the electro- and photoproduction processes, it makes
sense to introduce the cross-section of the subprocess $W/Z+p\to M+p$,
which has the form, 
\begin{eqnarray}
\frac{d\sigma_{W}}{dt} & = & \frac{G_{F}M_{W}^{2}}{\sqrt{2}}\frac{x_{Bj}^{2}\left|T_{M}\right|^{2}}{16\pi Q^{4}\left(1+\frac{4m_{N}^{2}x_{Bj}^{2}}{Q^{2}}\right)},\label{eq:XSec_Wp}\\
\frac{d\sigma_{Z}}{dt} & = & \frac{\sqrt{2}G_{F}M_{W}^{2}}{\cos^{2}\theta_{W}}\frac{x_{Bj}^{2}\left|T_{M}\right|^{2}}{16\pi Q^{4}\left(1+\frac{4m_{N}^{2}x_{Bj}^{2}}{Q^{2}}\right)}=\frac{\sqrt{2}G_{F}M_{Z}^{2}x_{Bj}^{2}\left|T_{M}\right|^{2}}{16\pi Q^{4}\left(1+\frac{4m_{N}^{2}x_{Bj}^{2}}{Q^{2}}\right)}
\end{eqnarray}
In neutrino experiments the target is unpolarized, so $\left|T_{M}\right|^{2}$
can be simplified to 
\begin{eqnarray}
\left|T_{M}\right|_{unp}^{2} & = & \frac{64\pi^{2}}{81}\frac{\alpha_{s}^{2}f_{M}^{2}}{Q^{2}(2-x_{Bj})^{2}}\left(\int dz\frac{\phi_{M}(z)}{z}\right)^{2}4\left[4\left(1-x_{Bj}\right)\left(\mathcal{H}_{M}\mathcal{H}_{M}^{*}+\tilde{\mathcal{H}}_{M}\tilde{\mathcal{H}}_{M}^{*}\right)-\frac{x_{Bj}^{2}t}{4m_{N}^{2}}\tilde{\mathcal{E}}_{M}\tilde{\mathcal{E}}_{M}^{*}\right.\label{eq:T2_unp}\\
 & - & \left.x_{Bj}^{2}\left(\mathcal{H}_{M}\mathcal{E}_{M}^{*}+\mathcal{E}_{M}\mathcal{H}_{M}^{*}+\tilde{\mathcal{H}}_{M}\tilde{\mathcal{E}}_{M}^{*}+\tilde{\mathcal{E}}_{M}\tilde{\mathcal{H}}_{M}^{*}\right)-\left(x_{Bj}^{2}+\left(2-x_{Bj}\right)^{2}\frac{t}{4m_{N}^{2}}\right)\mathcal{E}_{M}\mathcal{E}_{M}^{*}\right]\nonumber 
\end{eqnarray}

\section{GPD and DA parametrizations}

\label{sec:Parametrizations} As was mentioned in the introduction,
in the case of the DVMP a large part of the uncertainty comes from the
DAs of the produced mesons. In spite of many model-dependent estimates,
so far DAs are poorly known. Experimentally, only the distribution
amplitudes of the pions and $\eta$-meson have been challenged, and
even in this case the situation remains rather controversial. The
early experiments CELLO and CLEO~\cite{Gronberg:1997fj}, which studied
the small-$Q^{2}$ behavior of the form factor $F_{M\gamma\gamma}$,
found it to be consistent with the asymptotic form, $\phi_{as}(z)=6\sqrt{2}f_{M}z(1-z)$.
Later the BABAR collaboration~\cite{Aubert:2009mc} found a rapid
growth of the form factor $Q^{2}\left|F_{\pi\gamma\gamma}\left(Q^{2}\right)\right|^{2}$
in the large-$Q^{2}$ regime. This observation drew attention to this
problem and gave birth to speculations that the pion DA could be far
from the asymptotic shape~\cite{Polyakov:2009je} (see also a recent
review by Brodsky \emph{et. al.} in~\cite{Brodsky:2011xx,Brodsky:2011yv}).
However, the most recent data from BELLE~\cite{Uehara:2012ag} did
not confirm the rapid growth found by BABAR. As was found in~\cite{Pimikov:2012nm,Bakulev:2012nh},
the Gegenbauer expansion coefficients of the pion DA $\phi_{2;\pi}(z)$
are small and at most give a $10\%$ correction for the minus-first
moment, based on the fits of BELLE, CLEO and CELLO data. For kaons
there is no direct measurements of the DAs, it is expected however 
that the deviations from the pion DA are parametrically suppressed
by the quark mass $m_{s}/GeV$. Numerically this corresponds to a 10-20\%
deviation.

For this reason in what follows we assume all the Goldstone DAs to
have the asymptotic form, 
\[
\phi_{2;\{\pi,K,\eta\}}(z)\approx\phi_{as}(z)=6\sqrt{2}f_{M}z(1-z).
\]
For the decay couplings we use standard values $f_{\pi}\approx93$~MeV,
$f_{K}\approx113$~MeV, $f_{\eta}\approx f_{K}$.

More than a dozen of different parametrizations for GPDs have been
proposed so far~\cite{Diehl:2000xz,Goloskokov:2008ib,Radyushkin:1997ki,Kumericki:2009uq,Kumericki:2011rz,Guidal:2010de,Polyakov:2008aa,Polyakov:2002wz,Freund:2002qf}.
While we neither endorse nor refute any of them, for the sake of concreteness
we select the parametrization~\cite{Goloskokov:2006hr,Goloskokov:2007nt,Goloskokov:2008ib},
which succeeded to describe HERA~\cite{Aaron:2009xp} and JLAB~\cite{Goloskokov:2006hr,Goloskokov:2007nt,Goloskokov:2008ib}
data on electro- and photoproduction of different mesons, so it might
provide a reasonable description of $\nu$DVMP. The parametrization
is based on the Radyushkin's double distribution ansatz. It assumes
additivity of the valence and sea parts of the GPDs, 
\[
H(x,\xi,t)=H_{val}(x,\xi,t)+H_{sea}(x,\xi,t),
\]
which are defined as 
\begin{eqnarray*}
H_{val}^{q} & = & \int_{|\alpha|+|\beta|\le1}d\beta d\alpha\delta\left(\beta-x+\alpha\xi\right)\,\frac{3\theta(\beta)\left((1-|\beta|)^{2}-\alpha^{2}\right)}{4(1-|\beta|)^{3}}q_{val}(\beta)e^{\left(b_{i}-\alpha_{i}\ln|\beta|\right)t};\\
H_{sea}^{q} & = & \int_{|\alpha|+|\beta|\le1}d\beta d\alpha\delta\left(\beta-x+\alpha\xi\right)\,\frac{3\, sgn(\beta)\left((1-|\beta|)^{2}-\alpha^{2}\right)^{2}}{8(1-|\beta|)^{5}}q_{sea}(\beta)e^{\left(b_{i}-\alpha_{i}\ln|\beta|\right)t};
\end{eqnarray*}
and $q_{val}$ and $q_{sea}$ are the ordinary valence and sea components
of PDFs. The coefficients $b_{i}$, $\alpha_{i}$, as well as the parametrization
of the input PDFs $q(x),\,\Delta q(x)$ and pseudo-PDFs $e(x),\,\tilde{e}(x)$
(which correspond to the forward limit of the GPDs $E,\,\tilde{E}$)
are discussed in~\cite{Goloskokov:2006hr,Goloskokov:2007nt,Goloskokov:2008ib}.
The unpolarized PDFs $q(x)$ in the limited range $Q^{2}\lesssim40$~GeV$^{2}$
roughly coincide with the CTEQ PDFs. Notice that in this model the
sea is flavor symmetric for asymptotically large $Q^{2}$, 
\begin{equation}
H_{sea}^{u}=H_{sea}^{d}=\kappa\left(Q^{2}\right)H_{sea}^{s},\label{eq:SeaFlavourSymmetry}
\end{equation}
where 
\[
\kappa\left(Q^{2}\right)=1+\frac{0.68}{1+0.52\ln\left(Q^{2}/Q_{0}^{2}\right)},\quad Q_{0}^{2}=4\, GeV^{2}.
\]

The equality of the sea components of the light quarks in~(\ref{eq:SeaFlavourSymmetry})
should be considered only as a rough approximation, since in the forward
limit the inequality $\bar{d}\not=\bar{u}$ was firmly established
by the E866/NuSea experiment~\cite{Hawker:1998ty}. For this reason,
predictions done with this parametrization of GPDs for the $p\rightleftarrows n$
transitions in the region $x_{Bj}\in(0.1...0.3)$ might slightly underestimate
the data.

\section{Numerical results and discussion}

\label{sec:Results}

\begin{figure}
\includegraphics[scale=0.4]{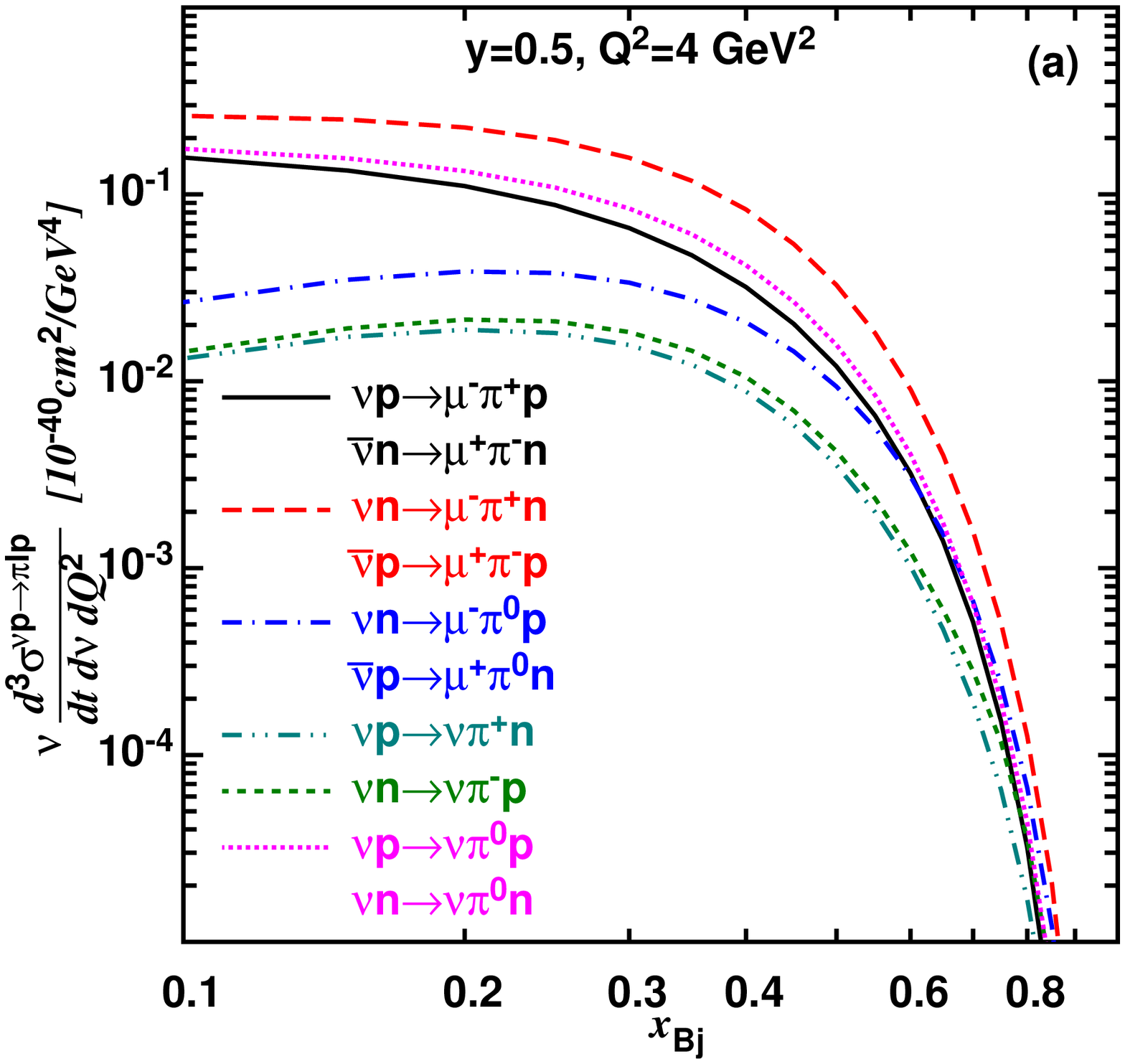}\qquad{}\includegraphics[scale=0.4]{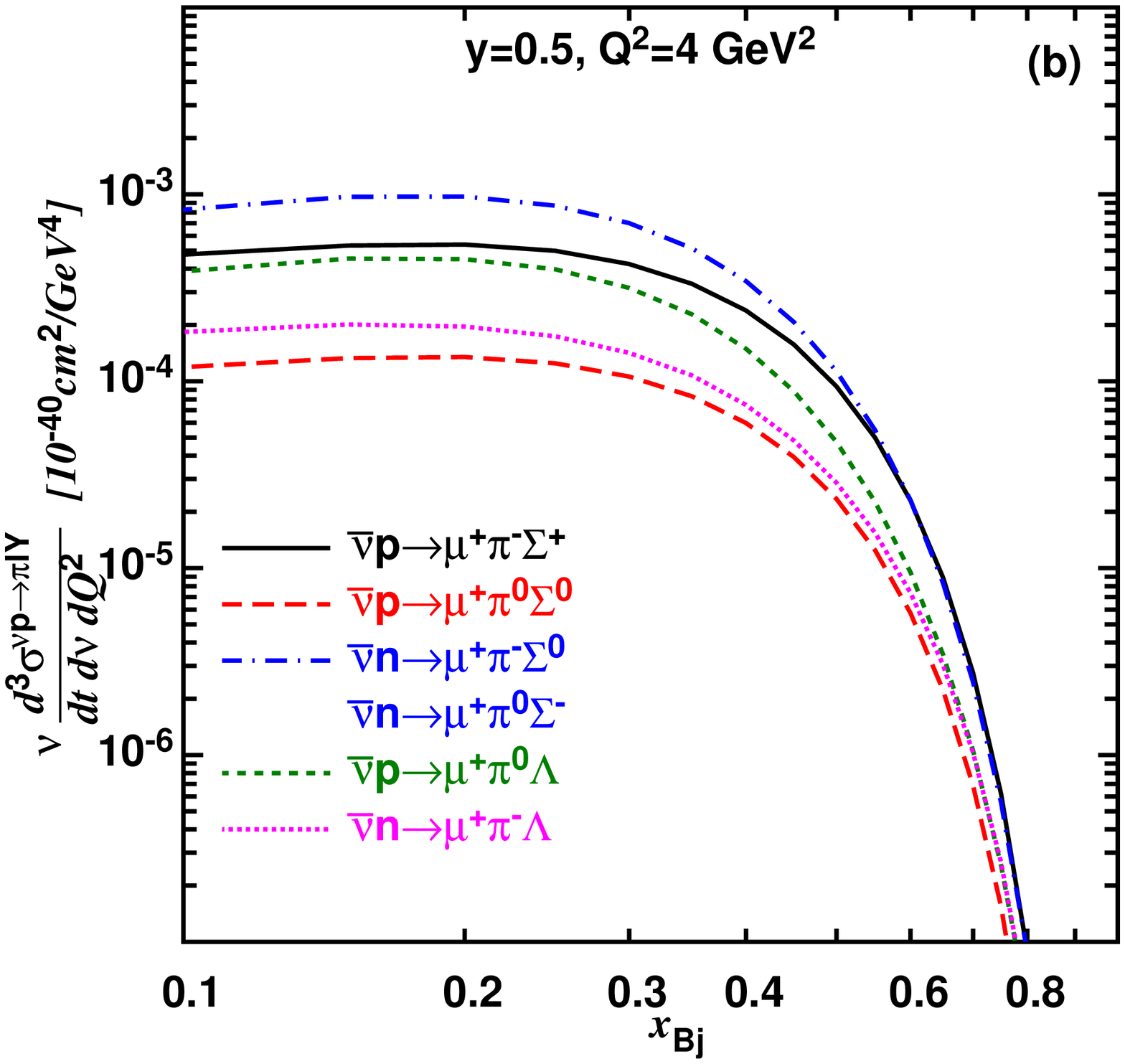}

\caption{\label{fig:DVMP-pions}(color online) Pion production on nucleons.
(a) Processes without strangeness production. (b) Processes with nucleon
to hyperon transition ($\Delta S=1$). Kinematics $t=t_{min}$ ($\Delta_{\perp}=0$)
is assumed.}
\end{figure}

In this section we perform numerical calculations of the cross-sections
of the processes listed in Table~\ref{tab:DVMP_amps} relying on
the GPDs presented in the previous section. The results for neutrino-production
of pions on nucleons are depicted in Figure~\ref{fig:DVMP-pions}.

In the left pane of the Figure~\ref{fig:DVMP-pions}, we grouped
the pion production processes without excitation of strangeness. The
results for the diagonal channels, $p\to p$ and $n\to n$ extend
to large $x_{Bj}$ our previous calculations \cite{Kopeliovich:2012tu}
for diffractive neutrino-production of pions performed in the dipole
approach, which assumes dominance of the sea. Differently from small-$x_{Bj}$
diffraction, in the valence quark region we found that the production
rate of $\pi^{+}$ on neutrons is about twice larger than on protons.
This results from the fact that the handbag diagram in the proton
probes the GPD $H_{d}$, whereas larger $H_{u}$ contributes via crossed
handbag; in the case of neutron they get swapped. At large $x_{Bj}\gtrsim0.6$
the corresponding cross-section is suppressed due to increase of $|t_{min}\left(x,Q^{2}\right)|$.

The off-diagonal processes with $p\rightleftarrows n$ transitions
are suppressed at small $x_{Bj}$ because they probe the GPD difference
$H_{u}-H_{d}$. In the small-$x_{Bj}$ regime ($x_{Bj}\lesssim0.1$)
the density of light sea quarks become equal, $\bar{d}\approx\bar{u}$,
and cancel. The valence quark PDFs and the invariant amplitude $T_{M}$
behave like $\sim1/\sqrt{x_{Bj}},$ so that the cross-section vanishes
as $\sim x_{Bj}$. This result agrees with the Regge phenomenology,
which predicts this cross-section to fall as%
\footnote{The pion intercept $\alpha_{\pi}(0)\approx0$, so the pion exchange
contribution is suppressed as $x_{Bj}^{2}$ %
} $\sim x_{Bj}^{2-2\alpha_{\rho}(0)}\approx x_{Bj}$~\cite{Guidal:1997hy}.
The cross-sections of the neutral $\pi^{0}$ production on the proton
and neutron (processes $\nu p\to\nu\pi^{0}p$ and $\nu n\to\nu\pi^{0}n$)
coincide under the assumption of $H$-dominance, however in the general
case they differ, with effects $\sim\tilde{H}_{u}-\tilde{H}_{d}$. Numerically
in the considered parametrization of GPDs these effects are of order
$1\,\%$, so the difference between the two curves is invisible in
the plot. A similar result holds for the processes $\nu p\to\nu\pi^{+}n$
and $\nu n\to\nu\pi^{-}p$: the corresponding cross-sections exactly
coincide under the assumption of $H$-dominance but differ in the general
case. As in neutral pion production, the difference is controlled
by a small $\sim\tilde{H}_{u}-\tilde{H}_{d}$, however, due to suppression
of the GPD $H$ in the small-$x_{Bj}$ region the effects proportional
$\sim\tilde{H}_{u}-\tilde{H}_{d}$ are relatively large, and the difference
between the two cross-sections becomes visible in the plot.

In the right pane we show the cross-sections of pion production with
nucleon to hyperon transition. These cross-sections are Cabibbo-suppressed
and hardly can be detected in the Minerva experiment. In contrast
to $p\rightleftarrows n$ processes, at small $x_{Bj}$ the sea contributes
to the difference $H^{u}-H^{s},\, H^{d}-H^{s}$.
First of all, the sea flavor asymmetry appears due to the presence of
proton nonperturbative Fock components, like $p\to K\Lambda$.
This asymmetry vanishes in the invariant amplitude at small $x_{Bj}$
as $x_{Bj}^{-\alpha_{K^{*}}}$, where the intercept of the $K^{*}$
Reggeon trajectory is $\alpha_{K^{*}}\approx0.25$. Correspondingly,
this contribution to the cross section is suppressed as $\sim x_{Bj}^{1.5}$.

At very small $x_{Bj}$ a more important source of flavor asymmetry
is the Pomeron itself. This is easy to understand within the dipole
approach, in which the sea-quark PDF probed for instance by a virtual
photon, corresponds to the transition $\gamma^{*}\to\bar{q}q$ and
interaction of the $\bar{q}q$ dipole with the target proton (in the
target rest frame). This cross section is different for $\bar{s}s$
and $\bar{u}u(\bar{d}d)$ dipoles due to so called aligned-jet configurations,
in which the dominant fraction of the dipole momentum is carried by
the quarks. Such a non-perturbative contribution to the PDF
is known to persist even at high $Q^{2}$ \cite{kp}. Such a flavor
asymmetry of the sea rises in the cross section as $x_{Bj}^{-2\epsilon(Q^{2})}$,
controlled by the value of the Pomeron intercept, $\epsilon(Q^{2})=\alpha_{\Pom}-1$,
which was found in the experiments at HERA \cite{epsilon} to grow
steeply with $Q^{2}$.

At medium-small values of $x_{Bj}$ in the transition region between
the two above regimes the interference of these two mechanisms, which
behaves as $x_{Bj}^{1-\alpha_{K^{*}}-\epsilon(Q^{2})}$, is also important.
This region of $x_{Bj}\gtrsim0.1$ is the domain of our main interest,
while small $x_{Bj}\ll1$ is beyond the scope of this paper. At $x_{Bj}\gtrsim0.2$,
the dominant contribution comes from the valence quarks.

\begin{figure}[tbh]
\includegraphics[scale=0.4]{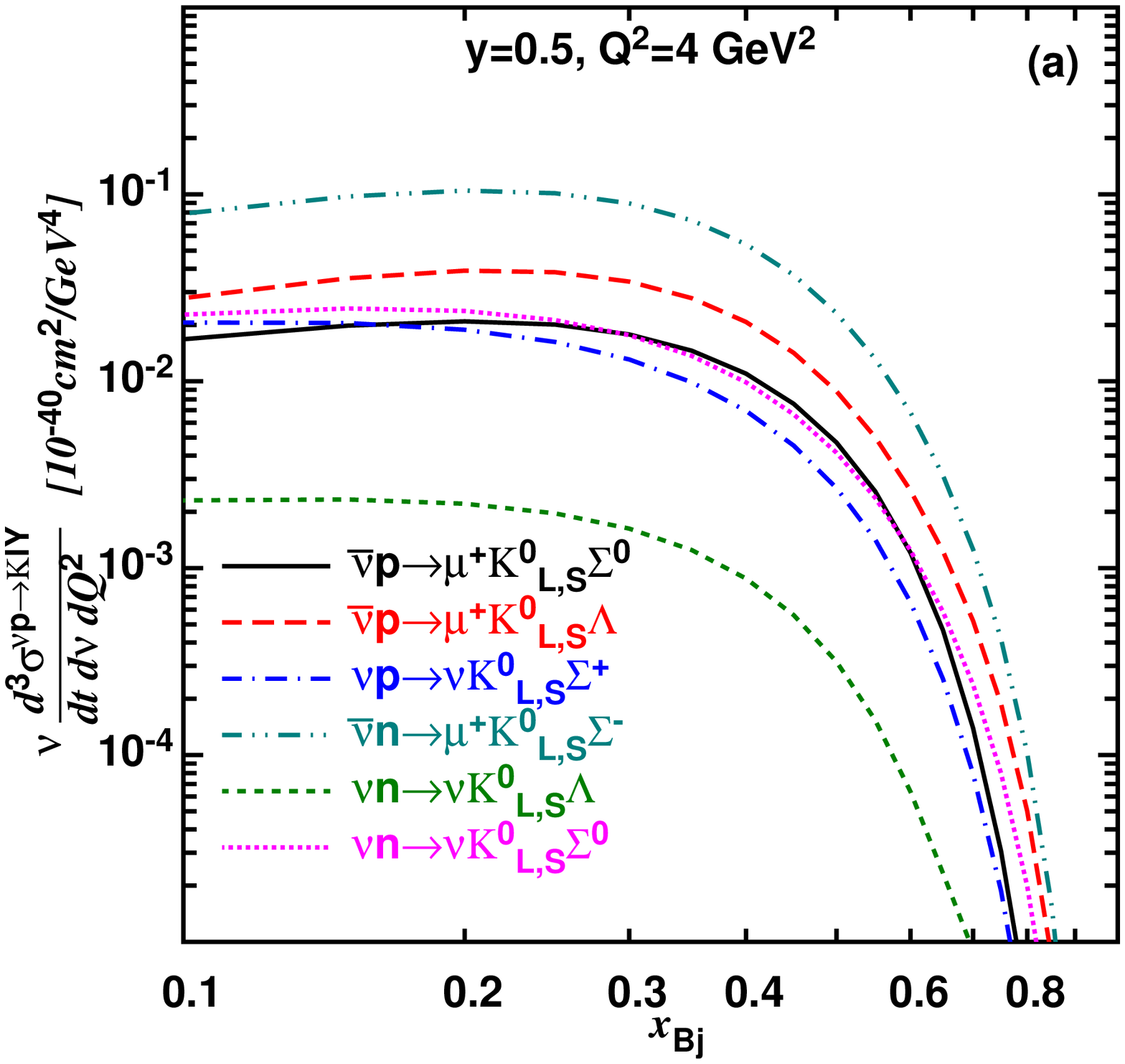}\qquad{}\includegraphics[scale=0.4]{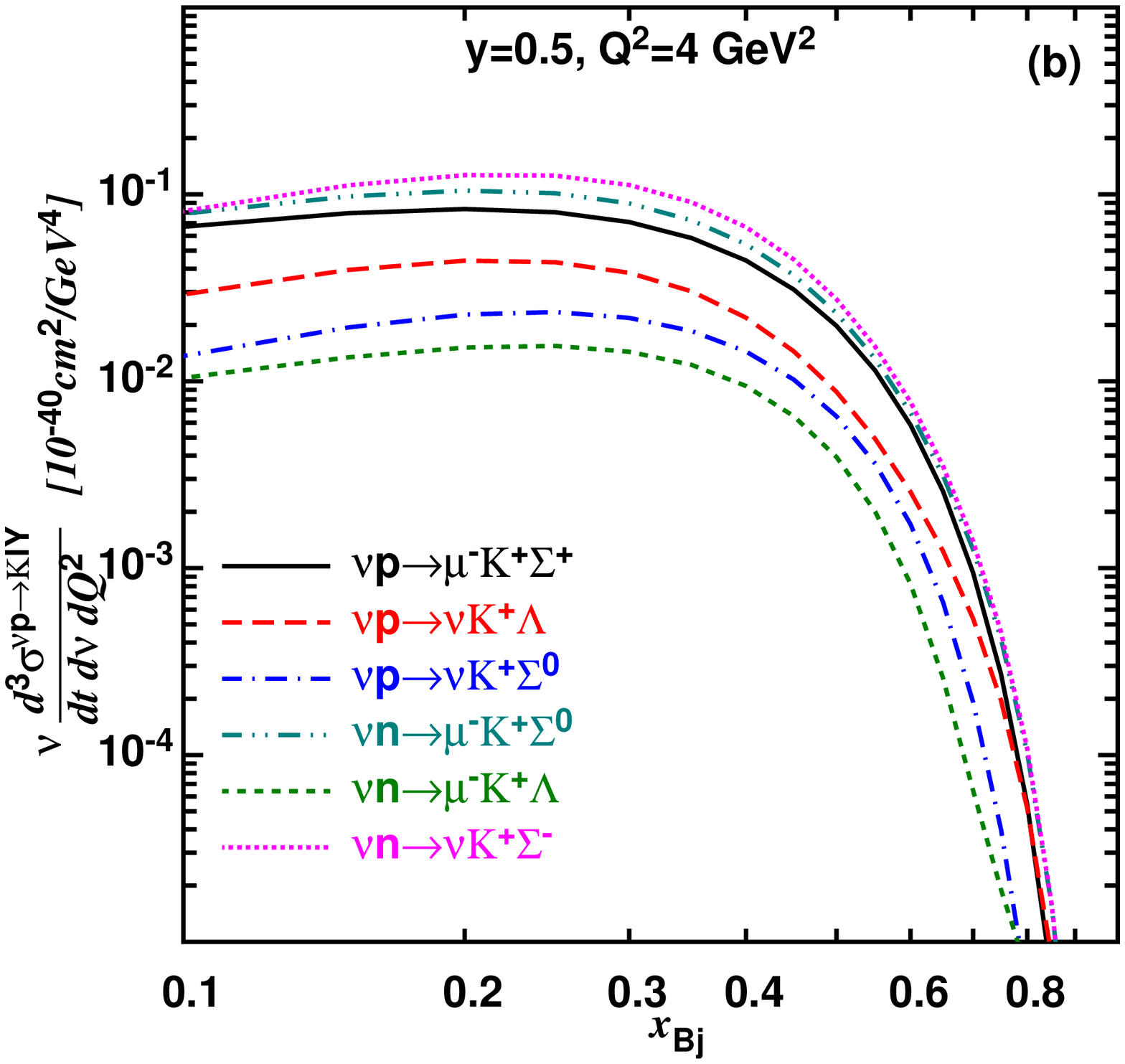}
\caption{\label{fig:DVMP-kaons-hyperon}(color online) Kaon production on the
nucleon with nucleon to hyperon transition ($\Delta S=0$). (a) Neutral
kaon production. (b) Charged kaon production. Kinematics $t=t_{min}$
($\Delta_{\perp}=0$) is assumed.}
\end{figure}

\begin{figure}[tbh]
\includegraphics[scale=0.4]{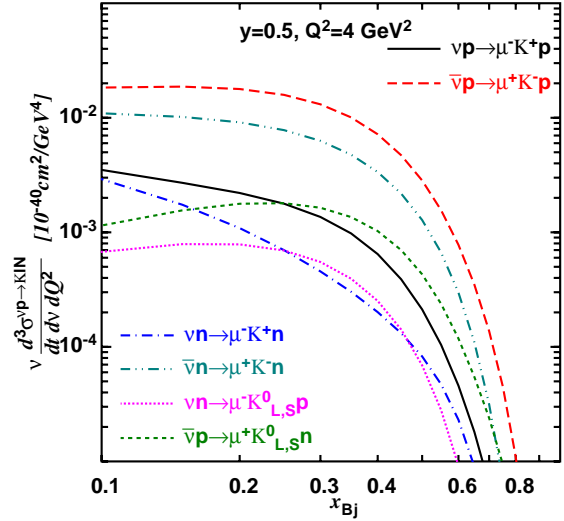} \caption{\label{fig:DVMP-kaons}(color online) Kaon production on the nucleon
without nucleon to hyperon transition ($\Delta S=1$). Kinematics
$t=t_{min}$ ($\Delta_{\perp}=0$) is assumed.}
\end{figure}

In Figures~\ref{fig:DVMP-kaons-hyperon},\ref{fig:DVMP-kaons}
we show the results for kaon production. Figure~\ref{fig:DVMP-kaons-hyperon}
presents the results for Cabibbo-allowed processes ($\Delta S=0$),
which involve a nucleon into the hyperon transition. For moderate values of
$x_{Bj}\gtrsim0.2$ the dominant contribution to these processes comes
from light quarks, and in the region of smaller $x_{Bj}$ from $s$-quarks.
In the left and right panes of Figure~\ref{fig:DVMP-kaons-hyperon},
we show the cross sections for neutral and charge kaon production
respectively. The order of magnitude of these cross sections is comparable
with that for pions, so should be easily measured at Minerva. However,
all these processes are suppressed at high energies (small-$x$ regime).
As was discussed in Section~\ref{sec:DVMP_Xsec}, there are two parametrization-independent
relations between the kaon-hyperon production cross-sections, $d\sigma_{\nu p\to\mu^{-}K^{+}\Sigma^{+}}=4\, d\sigma_{\bar{\nu}p\to\mu^{+}K^{0}\Sigma^{0}}$
and $d\sigma_{\bar{\nu}n\to\mu^{+}K_{L,S}\Sigma^{-}}=d\sigma_{\nu n\to\mu^{-}K^{+}\Sigma^{0}}.$
In Figure~\ref{fig:DVMP-kaons} one can see the cross-section
without hyperon formation. Such diagrams are Cabibbo suppressed (have
$\Delta S=1$) and have too small cross-sections, hardly detectable
in current experiments.

\begin{figure}
\includegraphics[scale=0.4]{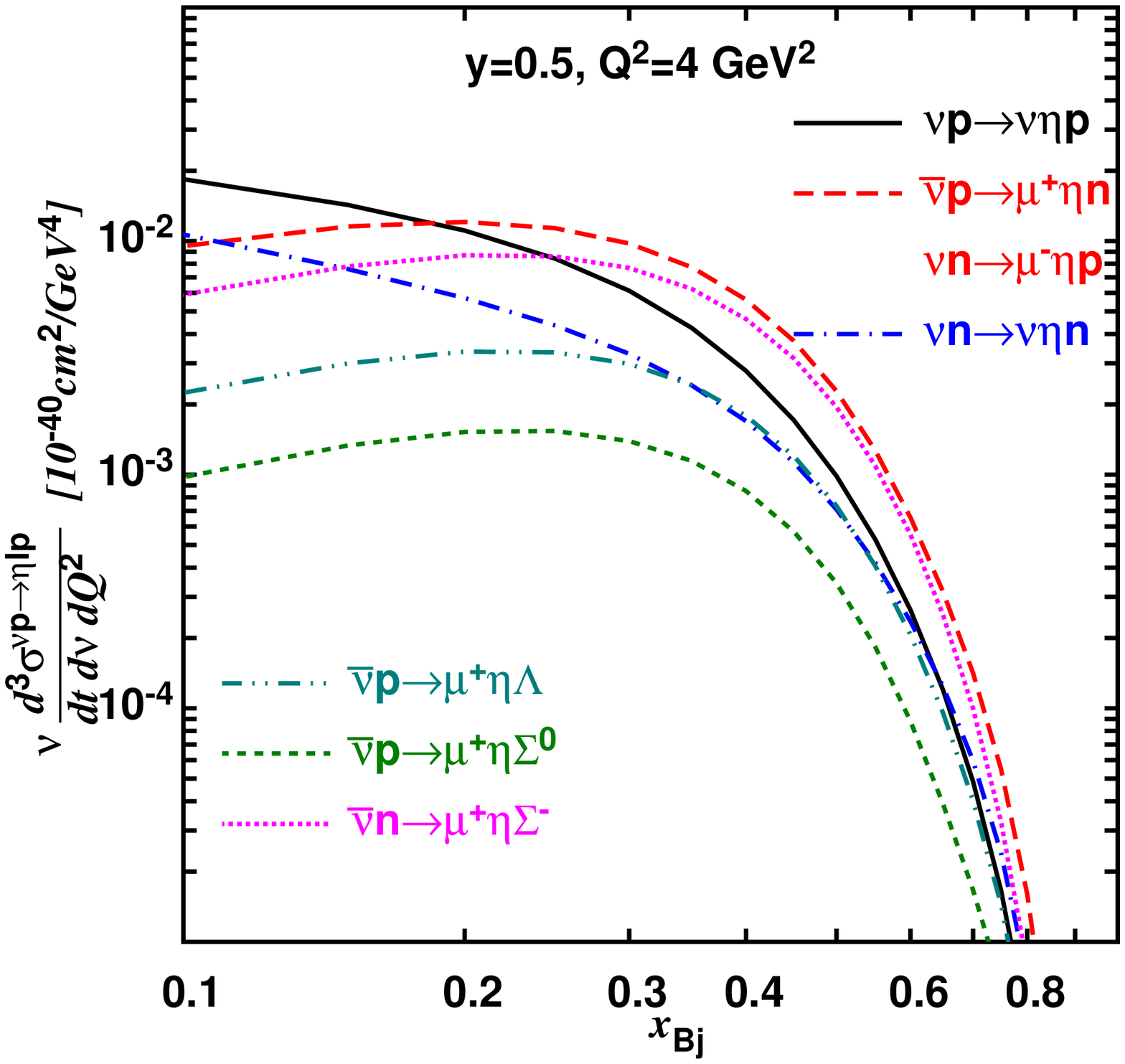} \caption{\label{fig:DVMP_eta}(color online) $\eta$-production on the nucleon.
In the upper right corner are the Cabibbo-allowed processes ($\Delta S=0$),
in the lower left corner are Cabibbo-suppressed processes ($\Delta S=1$).
Kinematics $t=t_{min}$ ($\Delta_{\perp}=0$) is assumed.}
\end{figure}

Figure~\ref{fig:DVMP_eta} demonstrates the results for $\eta$-production,
both with $\Delta S=0$ and $\Delta S=1$. The cross-sections for
all such processes at moderate values of $x_{Bj}$ are found to be
alike and to have similar dependences on $x_{Bj}$. Numerically, they
differ at most by an order of magnitude. At small $x_{Bj}$ however,
only the neutral current cross-sections $\nu p(n)\to\nu\eta p(n)$
survive, which depends on $x_{Bj}$ with the chosen parametrization
of GPDs as $\sim x_{Bj}^{0.2}$.

\begin{figure}
\includegraphics[scale=0.4]{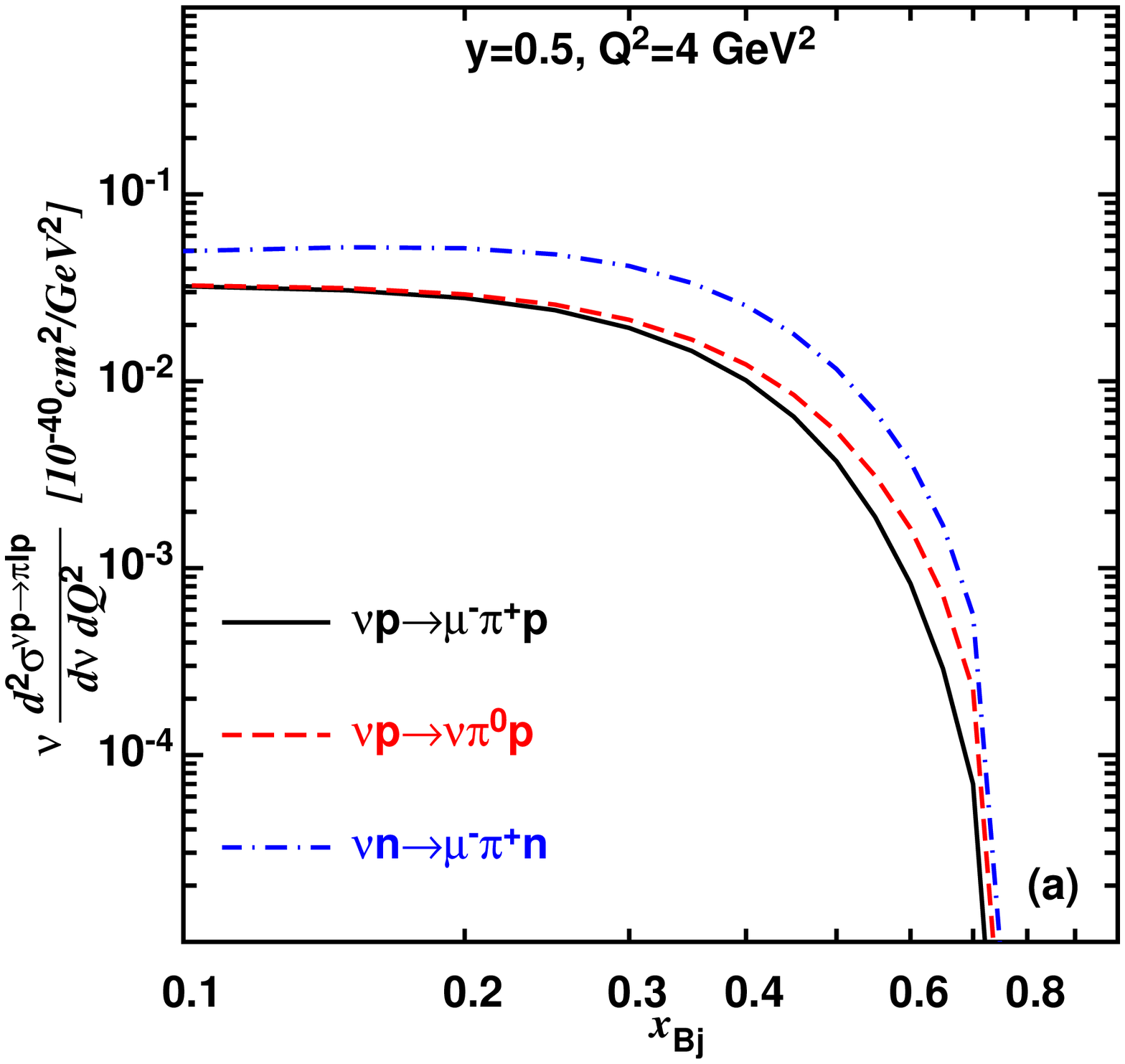}
\qquad{}\includegraphics[scale=0.4]{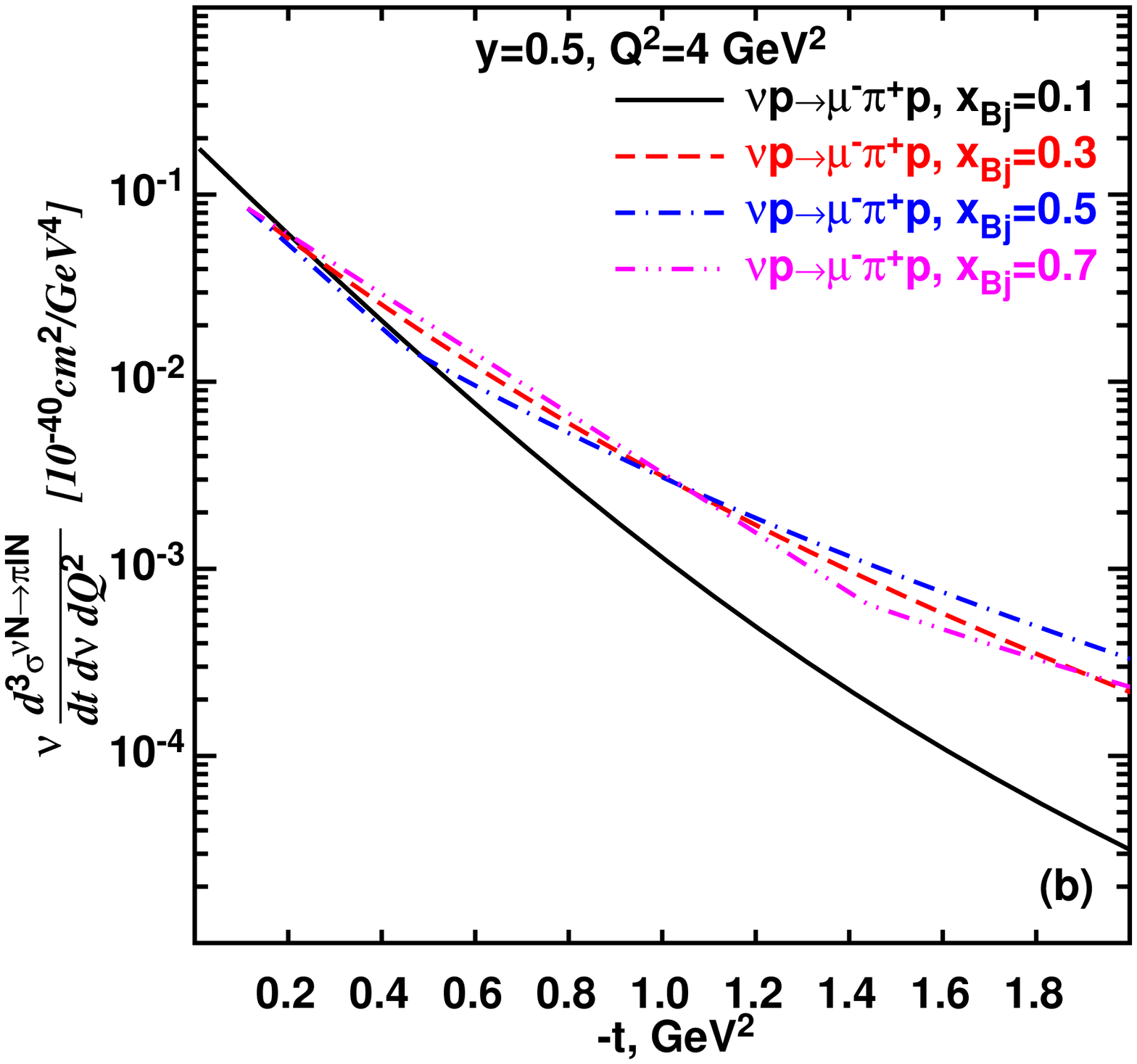}
\caption{\label{fig:t-dependence}(color online) (a) $t$-integrated two-fold
cross-section $d^{2}\sigma/d\nu\, dQ^{2}$. (b) $t$-dependence of
the differential cross-sections for selected processes.}
\end{figure}

The $t$-integrated cross section of pion production, calculated for
diagonal transitions and plotted in the left pane of Figure~\ref{fig:t-dependence}
demonstrates the features similar to the forward cross section depicted
in Figure~\ref{fig:DVMP-pions}. 

The cross section is steeply falling toward large $x_{Bj}$ due to
increasing $|t_{min}|$, whereas at small $x_{Bj}$ it behaves similarly
to the unintegrated cross section. Although in the valence region
the cross-sections may differ up to a factor of two, all diagonal
channels for charged and neutral currents unify at the same production
rate at small $x_{Bj}$, confirming the results of the dipole description
\cite{Kopeliovich:2012tu}.

\begin{figure}
\includegraphics[scale=0.4]{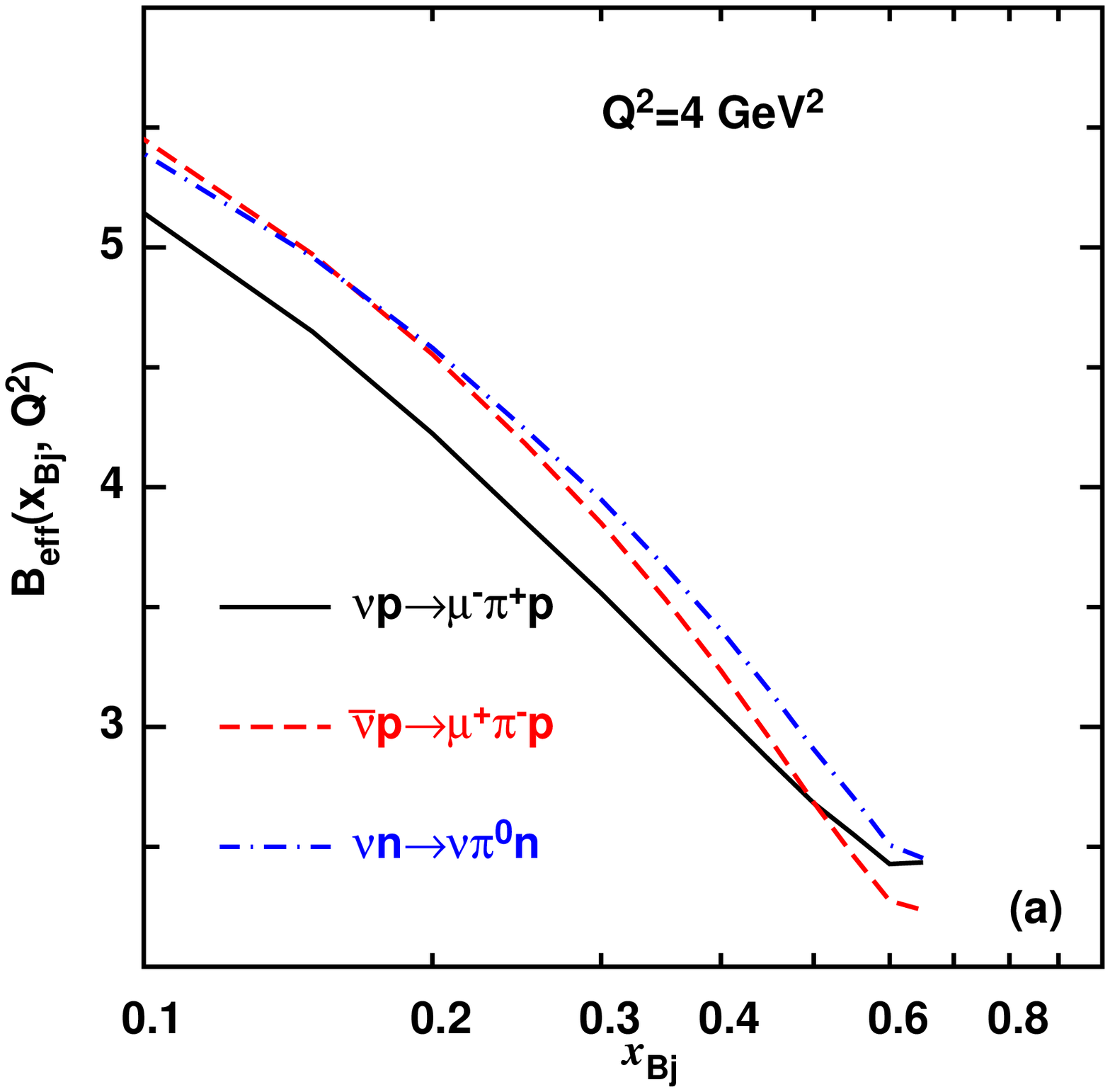}\qquad{}
\includegraphics[scale=0.4]{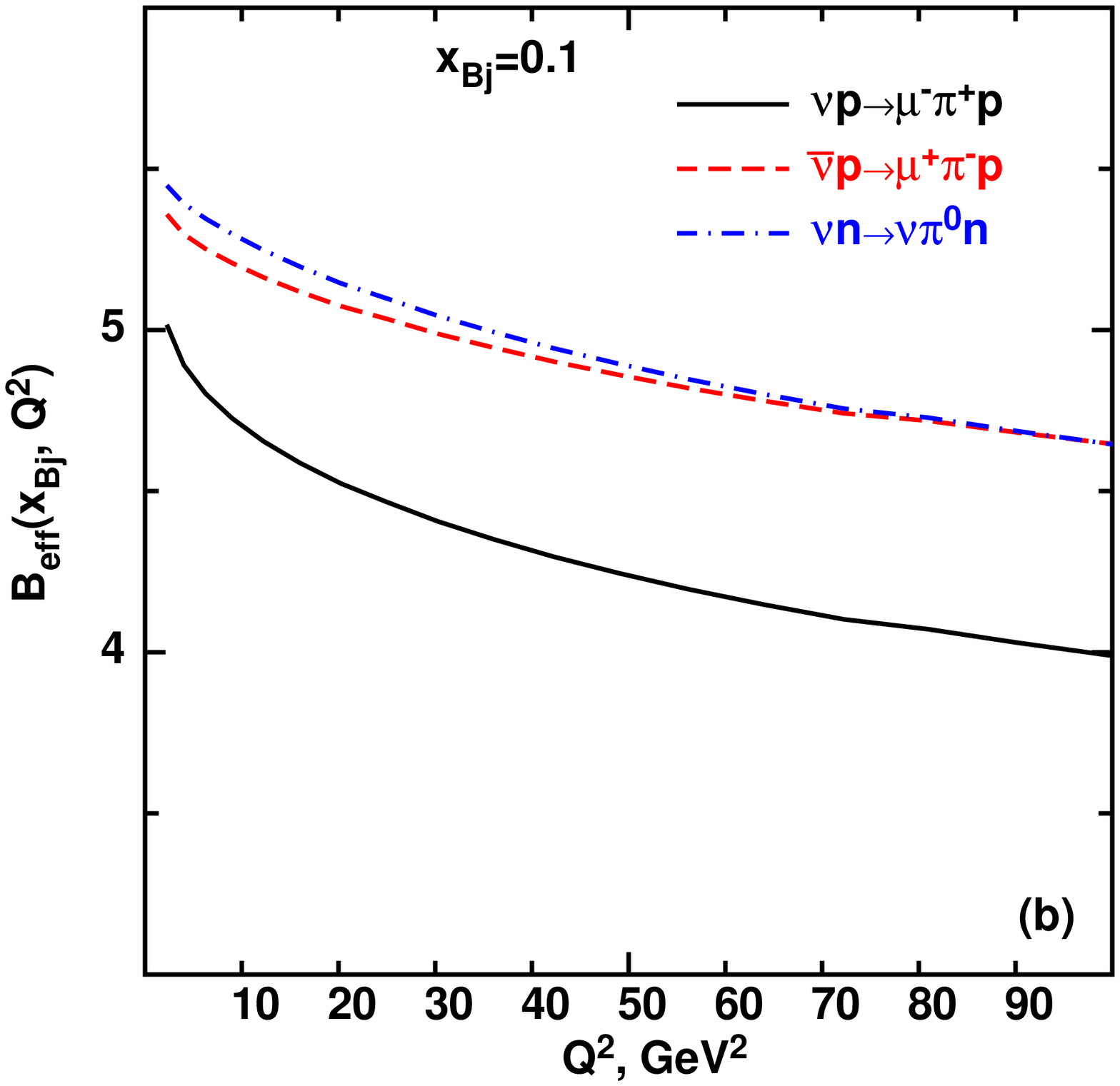}\caption{\label{fig:Beff}(color online) Effective slope $B_{eff}$ which controls
the $t$-dependence of the differential cross-section, as a function
of $x_{Bj}$ (left) and virtuality $Q^{2}$(right).}
\end{figure}

The $t$-dependence of the differential cross-section is controlled
by the underlying parametrization of GPDs we rely upon, and our results
for the differential cross section of neutrino-production of pions
are plotted in the right pane of Figure~\ref{fig:t-dependence}.
It can be roughly approximated by the exponential $t$-dependence
$d\sigma\sim\exp(B_{eff}t)$, with the slope 
\[
B_{eff}\equiv\frac{1}{\sigma_{W}}\left.\frac{d\sigma_{WN\to MN}}{dt}\right|_{t=0},
\]
which decreases with $x_{Bj}$ from about 6 down to 2~GeV$^{-2}$,
as is shown in the left pane of Figure~\ref{fig:Beff}. To a good
extent the calculated $x_{Bj}$-dependence of the slope is described
by $B(x)=B_{0}-\beta\ln(x),$ where the coefficient $\beta$ may vary
between $0.3$ ($\approx2\alpha'_{sea}(0)$) and $1.8$ $(=2\alpha'_{val}(0))$,
depending on the process and value of $Q^{2}$.

The $Q^{2}$ dependence of $B_{eff}$ depicted in the right pane of
Figure~\ref{fig:Beff} is rather mild, which is due to the weak dependence
of the shape of PDFs on $Q^{2}$ under DGLAP evolution.

\begin{figure}
\includegraphics[scale=0.4]{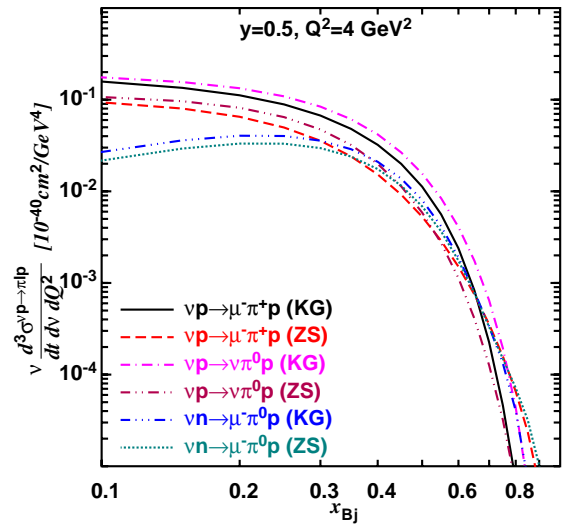}\caption{\label{fig:DVMP-KGvsZS}(color online) $\nu$DVMP cross-section for
certain processes. The abbreviations KG and ZS stand for the Kroll-Goloskokov
\cite{Goloskokov:2006hr,Goloskokov:2007nt,Goloskokov:2008ib} and
zero skewness (\ref{eq:ZS}) parametrizations.}
\end{figure}

Finally, in order to demonstrate explicitly the effect of skewness
we compare in Figure~\ref{fig:DVMP-KGvsZS} the cross-sections
of several processes calculated in the model \cite{Goloskokov:2006hr,Goloskokov:2007nt,Goloskokov:2008ib},
and with the simple zero-skewness parametrization 
\begin{equation}
H_{f}\left(x,\xi,t\right)\approx q_{_{f}}(x)F_{N}(t),\label{eq:ZS}
\end{equation}
where $q_{f}(x)$ is the parton distribution, and $F_{N}(t)$ is the
nucleon form factor. We see that the results of the two parametrizations
differ up to a factor of two.

\section{Summary}

We evaluated the cross-sections of deeply virtual meson production
for pions, kaons and eta-mesons in neutrino-nucleon interactions.
The production rate of the Cabibbo-allowed processes in the Bjorken
regime is found to be sufficiently large to be detected at 
the current level of statistics of neutrino experiments, in particular 
in the Minerva experiment at Fermilab, 
with accuracy, which allows to disentangle between different models
of GPDs. For this purpose we provided detailed information on the distributions 
of the production rate versus different variables, $x_{Bj}$ $t$, $Q^2$.
For further practical applications we provide a computational code.
We also evaluated the cross sections of Cabibbo-suppressed channels
($\Delta S=1$), but found them too weak to be detected by any of
forthcoming experiments.

\subsection*{Note added in proof}
After this manuscript was submitted, we learned that neutrino-production
of pions was also studied in~\cite{Goldstein:2009in} within the approach
proposed in~\cite{Ahmad:2008hp}. The central assumption
of~\cite{Goldstein:2009in} is that the pion coupling has a $\bar q\gamma_5
q$ structure, i.e. the dominant contribution comes from the subleading
twist pion DA $\phi_p$. This contribution represents a
$\mathcal{O}(1/Q^2)$
correction to the longitudinally polarized cross-sections discussed in this
paper, however it may be numerically important for the transversely
polarized current. Then, the amplitude gets contributions from
transversity GPDs $H_T$,
$E_T$, $\tilde H_T$, $\tilde E_T$. We did not include such corrections,
because
apart from the uncertainty in the chiral odd GPDs, this requires
modelling of the
poorly known twist-three pion DA $\phi_p$. No model-independent estimate
for this quantity is available so far, while model-dependent results vary
considerably~\cite{Zhong:2011jf,Ball:1998je}.

\section*{Acknowledgments}

This work was supported in part by Fondecyt (Chile) grants No. 1090291,
1100287 and 1120920.


\begin{thebibliography}{10}
  \bibitem{Kumericki:2009uq}K.~Kumericki and D.~Mueller, 
   Nucl.~Phys.~B \textbf{841}, 1 (2010) [arXiv:0904.0458 [hep-ph]].
  
  \bibitem{Boffi:2007yc}S.~Boffi and B.~Pasquini, Riv.~Nuovo Cim.~\textbf{30},
  387 (2007) [arXiv:0711.2625 [hep-ph]].
  
  \bibitem{Belitsky:2005qn} A.~V.~Belitsky and A.~V.~Radyushkin,
  Phys.\ Rept.\ \textbf{418}, 1 (2005) [arXiv:hep-ph/0504030].
  
  \bibitem{Diehl:2004cx}M.~Diehl, T.~Feldmann, R.~Jakob and P.~Kroll,
  Eur.~Phys.~J.~C \textbf{39}, 1 (2005) [hep-ph/0408173].
  
  \bibitem{Diehl:2003ny} M.~Diehl, Phys.\ Rept.\ \textbf{388}, 41
  (2003) [arXiv:hep-ph/0307382].
  
  \bibitem{Sabatie:2012pe}F.~Sabatie and H.~Moutarde, 
   PoS QNP \textbf{2012}, 016 (2012) [arXiv:1207.4655 [hep-ex]].
  
  \bibitem{Ji:1998xh} X.~D.~Ji and J.~Osborne, Phys.\ Rev.\ D
  \textbf{58} (1998) 094018 [arXiv:hep-ph/9801260].
  
  \bibitem{Collins:1998be} J.~C.~Collins and A.~Freund, Phys.\ Rev.\ D
  \textbf{59}, 074009 (1999).
  
  \bibitem{Guidal:2002kt}M.~Guidal and M.~Vanderhaeghen, Phys.~Rev.~Lett.~\textbf{90},
  012001 (2003) [hep-ph/0208275].
  
  \bibitem{Belitsky:2002tf}A.~V.~Belitsky and D.~Mueller, Phys.~Rev.~Lett.~
  \textbf{90}, 022001 (2003) [hep-ph/0210313].
  
  \bibitem{Belitsky:2003fj}A.~V.~Belitsky and D.~Mueller, Phys.~Rev.~D
  \textbf{68}, 116005 (2003) [hep-ph/0307369].
  
  \bibitem{Mueller:1998fv} D.~Mueller, D.~Robaschik, B.~Geyer, F.~M.~Dittes
  and J.~Horejsi, Fortsch.\ Phys.\ \textbf{42}, 101 (1994) [arXiv:hep-ph/9812448].
  
  \bibitem{Ji:1996nm} X.~D.~Ji, Phys.\ Rev.\ D \textbf{55}, 7114
  (1997).
  
  \bibitem{Ji:1998pc} X.~D.~Ji, J.\ Phys.\ G \textbf{24}, 1181
  (1998) [arXiv:hep-ph/9807358].
  
  \bibitem{Radyushkin:1996nd} A.~V.~Radyushkin, Phys.\ Lett.\ B
  \textbf{380}, 417 (1996) [arXiv:hep-ph/9604317].
  
  \bibitem{Radyushkin:1997ki} A.~V.~Radyushkin, Phys.\ Rev.\ D
  \textbf{56}, 5524 (1997).
  
  \bibitem{Radyushkin:2000uy} A.~V.~Radyushkin, arXiv:hep-ph/0101225.
  
  \bibitem{Collins:1996fb} J.~C.~Collins, L.~Frankfurt and M.~Strikman,
  Phys.\ Rev.\ D \textbf{56}, 2982 (1997).
  
  \bibitem{Brodsky:1994kf} S.~J.~Brodsky, L.~Frankfurt, J.~F.~Gunion,
  A.~H.~Mueller and M.~Strikman, Phys.\ Rev.\ D \textbf{50}, 3134
  (1994).
  
  \bibitem{Goeke:2001tz} K.~Goeke, M.~V.~Polyakov and M.~Vanderhaeghen,
  Prog.\ Part.\ Nucl.\ Phys.\ \textbf{47}, 401 (2001) [arXiv:hep-ph/0106012].
  
  \bibitem{Diehl:2000xz} M.~Diehl, T.~Feldmann, R.~Jakob and P.~Kroll,
  Nucl.\ Phys.\ B \textbf{596}, 33 (2001) [Erratum-ibid.\ B \textbf{605},
  647 (2001)] [arXiv:hep-ph/0009255].
  
  \bibitem{Belitsky:2001ns} A.~V.~Belitsky, D.~Mueller and A.~Kirchner,
  Nucl.\ Phys.\ B \textbf{629}, 323 (2002) [arXiv:hep-ph/0112108].
  
  \bibitem{Kubarovsky:2011zz}V.~Kubarovsky [CLAS Collaboration],
  Nucl.~Phys.~Proc.~Suppl.~\textbf{219-220}, 118 (2011).
  
  \bibitem{Vanderhaeghen:1998uc}M.~Vanderhaeghen, P.~A.~M.~Guichon
  and M.~Guidal, 
   Phys.~Rev.~Lett.~\textbf{80}, 5064 (1998).
  
  \bibitem{Mankiewicz:1998kg}L.~Mankiewicz, G.~Piller and A.~Radyushkin,
  Eur. Phys. J. C \textbf{10}, 307 (1999) [hep-ph/9812467].
  
  \bibitem{Ivanov:2007je}D.~Y.~Ivanov, arXiv:0712.3193 [hep-ph].
  
  \bibitem{GolecBiernat:1998js}K.~J.~Golec-Biernat and M.~Wusthoff,
   Phys.~Rev.~D \textbf{59}, 014017 (1998) [hep-ph/9807513].
  
  \bibitem{Hufner:2000jb} J.~Hufner, Yu.~P.~Ivanov, B.~Z.~Kopeliovich
  and A.~V.~Tarasov, 
   Phys.\ Rev.\ D \textbf{62} (2000) 094022 [arXiv:hep-ph/0007111].
  
  \bibitem{Kopeliovich:1999am} B.~Z.~Kopeliovich, A.~Schafer and
  A.~V.~Tarasov, 
   Phys.\ Rev.\ D \textbf{62} (2000) 054022 [arXiv:hep-ph/9908245].
  
  
  \bibitem{Kowalski:2003hm}H.~Kowalski and D. Teaney, Phys. Rev. D
  \textbf{68} (2003) 114005 [arXiv:hep-ph/0304189].
  
  \bibitem{Gronberg:1997fj}J.~Gronberg \emph{et al.} [CLEO Collaboration],
  Phys.~Rev. D \textbf{57}, 33 (1998) [hep-ex/9707031].
  
  \bibitem{Aubert:2009mc}B.~Aubert \emph{et al.} [BABAR Collaboration],
  Phys.~Rev.~D \textbf{80}, 052002 (2009) [arXiv:0905.4778 [hep-ex]].
  
  \bibitem{Uehara:2012ag}S.~Uehara \emph{et al.} [Belle Collaboration],
  arXiv:1205.3249 [hep-ex].
  
  \bibitem{Brodsky:2011xx}S.~J. Brodsky, F.~-G.~Cao and G.~F.~de
  Teramond, 
   Phys.~Rev.~D \textbf{84}, 075012 (2011) [arXiv:1105.3999 [hep-ph]].
  
  \bibitem{Brodsky:2011yv}S.~J.~Brodsky, F.~-G.~Cao and G.~F.~de
  Teramond, 
   Phys.~Rev.~D \textbf{84}, 033001 (2011) [arXiv:1104.3364 [hep-ph]].
  
  \bibitem{AlvarezRuso:2012fc}L.~Alvarez-Ruso, J.~Nieves, I.~R.~Simo,
  M.~Valverde and M.~J.~Vicente Vacas, 
   arXiv:1205.4863 [nucl-th].
  
  \bibitem{Alarcon:2011kh}J.~M.~Alarcon, J.~Martin Camalich, J.~A.~Oller
  and L.~Alvarez-Ruso, 
   Phys.~Rev.~C \textbf{83} (2011) 055205 [arXiv:1102.1537 [nucl-th]].
  
  \bibitem{Praet:2008yn}C.~Praet, O. Lalakulich, N.~Jachowicz and
  J.~Ryckebusch, 
   \textbf{79}, 044603 (2009) [arXiv:0804.2750 [nucl-th]].
  
  \bibitem{Paschos:2011ye}E.~A. Paschos and D.~Schalla, 
   Phys.~Rev.~D \textbf{84}, 013004 (2011) [arXiv:1102.4466 [hep-ph]].
  
  \bibitem{RafiAlam:2010kf}M.~Rafi Alam, I.~Ruiz Simo, M.~Sajjad
  Athar and M.~J.~Vicente Vacas, Phys.~Rev.~D \textbf{82}, 033001
  (2010) [arXiv:1004.5484 [hep-ph]].
  
  \bibitem{Drakoulakos:2004gn}D.~Drakoulakos \emph{et al.} [Minerva
  Collaboration], 
   hep-ex/0405002.
  
  \bibitem{Gallardo:1996aa}J. C. Gallardo, R.~B. Palmer, A.~V.~Tollestrup,
  A. M.~Sessler, A.~N.~Skrinsky, C.~Ankenbrandt, S.~Geer and J.~Griffin
  \emph{et al.}, 
eConf C \textbf{960625} (1996) R4.
  
  \bibitem{Ankenbrandt:1999as}C.~M.~Ankenbrandt, M.~Atac, B.~Autin,
  V.~I.~Balbekov, V.~D.~Barger, O.~Benary, J.~S.~Berg and M.~S.~Berger
  \emph{et al.}, 
   Phys.~Rev.~ST Accel.~Beams \textbf{2} (1999) 081001 [physics/9901022].
  
  \bibitem{Alsharoa:2002wu}M.~M.~Alsharoa \emph{et al.} [Muon Collider/Neutrino
  Factory Collaboration], 
   Phys.~Rev.~ST Accel.~Beams \textbf{6} (2003) 081001 [hep-ex/0207031].
  
  \bibitem{Psaker:2006gj}A. Psaker, W.~Melnitchouk and A.~V. Radyushkin,
  Phys.~Rev.~D \textbf{75}, 054001 (2007) [hep-ph/0612269].
  
  \bibitem{Frankfurt:1999fp}L.~L.~Frankfurt, P.~V.~Pobylitsa, M.~V.~Polyakov
  and M.~Strikman, 
   Phys.~Rev.~D \textbf{60} (1999) 014010 [hep-ph/9901429].
  
  \bibitem{Ivanov:2004zv}D.~Y.~Ivanov, L.~Szymanowski and G.~Krasnikov,
  JETP Lett.~ \textbf{80}, 226 (2004) [Pisma Zh.~Eksp.~Teor.~Fiz.~\textbf{80},
  255 (2004)] [hep-ph/0407207].
  
  \bibitem{Diehl:2007hd}M.~Diehl and W.~Kugler, Eur.~Phys.~J.~C
  \textbf{52}, 933 (2007) [arXiv:0708.1121 [hep-ph]].
  
  \bibitem{Guidal:1997hy}M.~Guidal, J.~M.~Laget and M.~Vanderhaeghen,
  Nucl.~Phys.~A \textbf{627} (1997) 645.
  
  \bibitem{Li:2004gu}D.~-M.~Li, B.~Ma, Y.~-X.~Li, Q.~-K.~Yao
  and H.~Yu, Eur.~Phys.~J.~C \textbf{37}, 323 (2004) [hep-ph/0408214].
  
  \bibitem{Polyakov:2009je}M.~V. Polyakov, 
   JETP Lett.~\textbf{90}, 228 (2009) [arXiv:0906.0538 [hep-ph]].
  
  \bibitem{Pimikov:2012nm}A.~V.~Pimikov, A.~P.~Bakulev, S.~V.~Mikhailov
  and N.~G.~Stefanis, arXiv:1208.4754 [hep-ph].
  
  \bibitem{Bakulev:2012nh}A.~P.~Bakulev, S.~V.~Mikhailov, A.~V.~Pimikov
  and N.~G.~Stefanis, Phys.~Rev.~D \textbf{86} (2012) 031501 [arXiv:1205.3770
  [hep-ph]].
  
  \bibitem{Goloskokov:2008ib}S.~V. Goloskokov and P.~Kroll, Eur.~Phys.~J.~C
  \textbf{59} (2009) 809 [arXiv:0809.4126 [hep-ph]].
  
  \bibitem{Kumericki:2011rz}K.~Kumericki, D.~Muller and A.~Schafer,
  JHEP \textbf{1107}, 073 (2011) [arXiv:1106.2808 [hep-ph]].
  
  \bibitem{Guidal:2010de}M.~Guidal, 
   Phys.~Lett.~B \textbf{693}, 17 (2010) [arXiv:1005.4922 [hep-ph]].
  
  \bibitem{Polyakov:2008aa}M.~V.~Polyakov and K.~M.~Semenov-Tian-Shansky,
  Eur.~Phys.~J.~A \textbf{40}, 181 (2009) [arXiv:0811.2901 [hep-ph]].
  
  \bibitem{Polyakov:2002wz}M.~V.~Polyakov and A.~G.~Shuvaev, 
   hep-ph/0207153.
  
  \bibitem{Freund:2002qf}A.~Freund, M.~McDermott and M.~Strikman,
  Phys.~Rev.~D \textbf{67}, 036001 (2003) [hep-ph/0208160].
  
  \bibitem{Goloskokov:2006hr}S.~V.~Goloskokov and P.~Kroll, 
   Eur.~Phys.~J.~C \textbf{50}, 829 (2007) [hep-ph/0611290].
  
  \bibitem{Goloskokov:2007nt}S.~V.~Goloskokov and P.~Kroll, 
   Eur.~Phys.~J.~C \textbf{53}, 367 (2008) [arXiv:0708.3569 [hep-ph]].
  
  \bibitem{Aaron:2009xp}F.~D.~Aaron \emph{et al.} [H1 Collaboration],
  JHEP \textbf{1005} (2010) 032 [arXiv:0910.5831 [hep-ex]].
  
  \bibitem{Hawker:1998ty}E.~A.~Hawker \emph{et al.} [FNAL E866/NuSea
  Collaboration], Phys.~Rev.~Lett.~\textbf{80} (1998) 3715 [hep-ex/9803011].
  
  \bibitem{Kopeliovich:2012tu}B.~Z.~Kopeliovich, I.~Schmidt and
  M.~Siddikov, Phys.~Rev.~D \textbf{85}, 073003 (2012) [arXiv:1201.4053
  [hep-ph]].
  
  \bibitem{kp} B.~Kopeliovich and B.~Povh, Z.\ Phys.\ A \textbf{356},
  467 (1997) [nucl-th/9607035].
  
  \bibitem{epsilon}J. Breitweg et al. (ZEUS Collaboration), Eur. Phys.  J. C \textbf{7}, 609 (1999). 

\bibitem{Goldstein:2009in} 
  G.~R.~Goldstein, O.~G.~Hernandez, S.~Liuti and T.~McAskill,
  AIP Conf.\ Proc.\  {\bf 1222}, 248 (2010)
  [arXiv:0911.0455 [hep-ph]].
\bibitem{Ahmad:2008hp}
  S.~Ahmad, G.~R.~Goldstein and S.~Liuti,
  Phys.\ Rev.\ D {\bf 79} (2009) 054014
  [arXiv:0805.3568 [hep-ph]].
\bibitem{Zhong:2011jf}
  T.~Zhong, X.~-G.~Wu, J.~-W.~Zhang, Y.~-Q.~Tang and Z.~-Y.~Fang,
  Phys.\ Rev.\ D {\bf 83} (2011) 036002
  [arXiv:1101.3592 [hep-ph]].
\bibitem{Ball:1998je}
  P.~Ball,
  JHEP {\bf 9901} (1999) 010
  [hep-ph/9812375].


\end{thebibliography}
  \end{document}